\documentclass{aa}  

\usepackage{graphicx}
\usepackage{txfonts}

\usepackage{hyperref}
\hypersetup{colorlinks,citecolor=blue}
\newcommand{\mjp}{miniJPAS}

\newcommand{\Ha}{\ifmmode \mathrm{H}\alpha \else H$\alpha$\fi\xspace}
\newcommand{\Hb}{\ifmmode \mathrm{H}\beta \else H$\beta$\fi\xspace}
\newcommand{\nii}{\ifmmode [\mathrm{N}\,\textsc{ii}] \else [N~{\scshape ii}]\fi\xspace}
\newcommand{\oiii}{\ifmmode [\rm{O}\,\textsc{iii}] \else [O\,{\sc iii}]\fi\xspace}
\newcommand{\oii}{\ifmmode [\rm{O}\,\textsc{ii}] \else [O\,{\sc ii}]\fi\xspace}
\newcommand{\sii}{\ifmmode [\mathrm{S}\,\textsc{ii}] \else [S~{\scshape ii}]\fi\xspace}

\begin{document}

\title{J-PAS: Measuring emission lines with artificial neural networks}

\author{G. Martínez-Solaeche \inst{\ref{1}} \and 
R. M. González Delgado\inst{\ref{1}} \and
R. García-Benito \inst{\ref{1}} \and  
A. de Amorim \inst{\ref{2}} \and 
E. Pérez \inst{\ref{1}} \and 
J. E. Rodríguez-Martín  \inst{\ref{1}} \and   
L.~A.~D\'iaz-Garc\'ia   \inst{\ref{1},\ref{14}} \and 
R. Cid Fernandes  \inst{\ref{2}} \and
C.~L\'opez-Sanjuan\inst{\ref{3}} \and
S.~Bonoli\inst{\ref{3},}\inst{\ref{4},}\inst{\ref{5}} \and
A.~J.~Cenarro\inst{\ref{3}} \and
R.~A.~Dupke\inst{\ref{6},}\inst{\ref{11},}\inst{\ref{12}} \and
A.~Mar\'in-Franch\inst{\ref{3}} \and
J.~Varela\inst{\ref{3}} \and
H.~V\'azquez~Rami\'o\inst{\ref{3}} \and
L.~R.~Abramo\inst{\ref{7}} \and
D.~Crist\'obal-Hornillos\inst{\ref{3}} \and
M.~Moles\inst{\ref{3},}\inst{\ref{1}} \and
J.~Alcaniz\inst{\ref{6},}\inst{\ref{8}} \and
P.O.~Baqui\inst{\ref{15}} \and
N.~Benitez\inst{\ref{1}} \and
S.~Carneiro\inst{\ref{9}} \and
 A.~Cortesi\inst{\ref{16}} \and
A.~Ederoclite\inst{\ref{10}} \and
V.~Marra\inst{\ref{15}} \and
C.~Mendes~de~Oliveira\inst{\ref{10}} \and
L.~Sodr\'e~Jr.\inst{\ref{10}} \and
J. M.~V\'ilchez \inst{\ref{1}} \and
K.~Taylor\inst{\ref{13}} \and
JPAS collaboration
}

\institute{Instituto de Astrofísica de Andalucía (CSIC), PO Box 3004, 18080 Granada, Spain (\email{gimarso@iaa.es}) \label{1} \and
Departamento de F\'{\i}sica, Universidade Federal de Santa Catarina, P.O. Box 476, 88040-900, Florian\'opolis, SC, Brazil \label{2} \and
Centro de Estudios de F\'isica del Cosmos de Arag\'on (CEFCA), Unidad Asociada al CSIC, Plaza San Juan, 1, E-44001, Teruel, Spain  \label{3} \and
Donostia International Physics Center (DIPC),  Manuel Lardizabal Ibilbidea, 4, San Sebasti\'an, Spain \label{4} \and
Ikerbasque, Basque Foundation for Science, E-48013 Bilbao, Spain \label{5} \and 
Observat\'orio Nacional, Minist\'erio da Ciencia, Tecnologia, Inovaç\~ao e Comunicaç\~oes, Rua General Jos\'e Cristino, 77, S\~ao Crist\'ov\~ao, 20921-400, Rio de Janeiro, Brazil \label{6} \and
Instituto de F\'isica, Universidade de S\~ao Paulo, Rua do Mat\~ao 1371, CEP 05508-090,  S\~ao Paulo, Brazil \label{7}\and
Departamento de F\'isica, Universidade Federal do Rio Grande do Norte, 59072-970, Natal, RN, Brazil \label{8}\and
Instituto de F\'isica, Universidade Federal da Bahia, 40210-340, Salvador, BA, Brazil \label{9} \and
Departamento de Astronomia, Instituto de Astronomia, Geofísica e Ciências Atmosf\'ericas, Universidade de São Paulo, São Paulo, Brazil \label{10} \and
Department of Astronomy, University of Michigan, 311West Hall, 1085 South University Ave., Ann Arbor, USA \label{11} \and
Department of Physics and Astronomy, University of Alabama, Box 870324, Tuscaloosa, AL, USA \label{12} \and
Instruments4 \label{13}
\and
Academia Sinica Institute of Astronomy \& Astrophysics (ASIAA), 11F of Astronomy-Mathematics Building, AS/NTU, No.~1, Sect. 4, Roosevelt Road, Taipei 10617, Taiwan \label{14} 
\and
 N\'ucleo de Astrof\'usica e Cosmologia (Cosmo-ufes) \& Departamento de F\'isica, Universidade Federal do Esp\'irito Santo, 29075-910, Vit\'oria, ES, Brazil
 \label{15}
 \and
 Observat\'orio do Valongo, Universidade Federal do Rio de Janeiro,
20080-090, Rio de Janeiro, RJ, Brazil
\label{16}}
\date{Received: 10-08-2020 / 
Accepted: 29/12/2020} 
  

   \abstract{\small{J-PAS will observe $8000$~deg$^2$ of the northern sky in the upcoming years with 56 photometric bands. J-PAS is a very suit survey for the detection of nebular emission objects. This paper presents a new method based on Artificial Neural Networks (ANNs) to measure and detect emission lines in galaxies up to $z = 0.35$. These lines are essential diagnostics to understand the evolution of galaxies through cosmic time. We trained and tested ANNs with synthetic J-PAS photometry from CALIFA, MaNGA, and SDSS spectra. We carry out two tasks: firstly, we cluster galaxies in two groups according to the values of the equivalent width (EW) of \Ha, \Hb, \nii, and \oiii lines measured in the spectra. Then, we train an ANN to assign to each galaxy a group. We are able to classify them with the uncertainties typical of the photometric redshift measurable in J-PAS. Secondly, we utilize another ANN to determine the values of those EWs. Subsequently, we obtain the \nii/\Ha,  \oiii/\Hb, and \ion{O}{3}\ion{N}{2} ratios recovering the BPT diagram (\oiii/\Hb vs \nii/\Ha). We study the performance of the ANN in two training samples: one is only composed of synthetic J-PAS photo-spectra (J-spectra) from MaNGA and CALIFA (CALMa set) and the other one is composed of SDSS galaxies. We can reproduce properly the main sequence of star forming galaxies from the determination of the EWs. With the CALMa training set we reach a precision of 0.093 and 0.081 dex for the \nii/\Ha and  \oiii/\Hb ratios in the SDSS testing sample. Nevertheless, we find an underestimation of those ratios at high values in galaxies hosting an active galactic nuclei. We also show the importance of the dataset used for both training and testing the model. ANNs are extremely useful to overcome the limitations previously expected concerning the detection and measurements of the emission lines in surveys like J-PAS. We show the capability of the method to measure a EW of 10 \AA $ $ in \Ha, \Hb, \nii and \oiii lines with a signal-to-noise ratio (SN) in the photometry of 5, 1.5, 3.5, and 10  respectively. Finally, we compare the properties of emission lines in galaxies observed with miniJPAS and SDSS. Despite of the limitation of such a comparison, we find a remarkable correlation in their EWs.}}
  \keywords{galaxies: evolution – surveys – techniques: photometric – methods: data analysis}
  
   \maketitle
%

\section{Introduction}
The study of the formation and evolution of galaxies through cosmic time has been addressed in the last decades by understanding how their physical properties leave footprints in the spectral energy distribution \citep[see e.g.][and references therein]{2019A&A...631A.158D}. Both the analysis of the light coming from stars and the ionized interstellar gas can be converted by well-known recipes to physical quantities such as the stellar mass, star formation rate (SFR), dust attenuation, luminosity-age, gas-phase metallicity or can unveil the main ionization mechanism responsible for the optical emission lines we observe in the spectrum \citep[for some of the most recent reviews on these topics, see][]{2013ARA&A..51..393C,2014ARA&A..52..415M,2019ARA&A..57..511K}.
\\\\The most massive and youngest stars within galaxies are responsible for the ultraviolet emission in the spectrum, but many times the presence of dust grains does not allow ultraviolet photons to travel freely through the interstellar medium and consequently makes it difficult to constrain the SFR from the blue part of the spectrum alone. However, those stars can actually ionize the surrounding interstellar gas. Very rapidly, hydrogen atoms recombine leaving tracks in form of emission lines at a particular wavelength in the spectrum. The Balmer series places \Ha  at $6562.8$ \AA, hence it is less affected by dust extinction and an excellent tracer to measure SFRs up to $z\sim0.4$ in the optical range \citep{2015A&A...584A..87C}. 
\\\\ Other lines, such as the forbidden \oiii$\lambda \lambda 4959, 5007$  \AA $ $ and \nii$\lambda \lambda 6548, 6584 $  \AA $ $ doublets\footnote{In the remaining of this paper, \oiii$\lambda 5007$  and  \nii$\lambda6584$ will be denoted \oiii and \nii, respectively.}, are sensitive to the gas-phase metallicity, which is ideal for investigating the metal enrichment of gas throughout cosmic time \citep{2019A&ARv..27....3M}. The\nii/\Ha and \oiii/\Hb ratios among others are used to construct the so-called BPT diagrams \citep{1981PASP...93....5B},  which distinguish galaxies where the gas has 
been ionized due to the presence of an active galactic nuclei (AGN) from those where the main ionization mechanism comes from high rates of star formation in the galaxy or shock ionized gas regions.
\\ \\ Even though spectroscopic surveys revolutionized astronomy in many fields, they provide a limited picture of the universe in many senses. Both Multi-Object Spectroscopy and integral fields units (IFUs) surveys are partially biased due to pre-selected samples where some properties such as fluxes, redshift or galaxy-size are limited to a certain range. Some of these issues can partially be solved with narrow band photometric surveys.  Although they have been historically limited to few filters, they can act as low-resolution spectrographs and they are able to map the sky quickly and deeply; therefore, giving a more comprehensive snapshot of the universe. Needless to say, some astrophysical analyses will always require the highest possible spectral resolution to fully exploit all the information encoded in the spectrum. 
\\ \\Maybe one of the most competitive astrophysical surveys designed to overcome the weakness of photometry and spectrography, halfway between them, is the Javalambre-Physics of the Accelerating Universe \citep[J-PAS,][]{2014arXiv1403.5237B}. It will sample the optical spectrum with 56 narrow-band filters for hundreds of millions of galaxies and stars over  $\sim8000$ deg$^2$. This is equivalent to a resolving power of $R \sim 50$ (J-spectrum hereafter). Initially thought to explore the origin and nature of the dark energy in the universe, J-PAS is also ideal for galaxy evolution studies and to detect emission line objects \citep{2020arXiv200701910B}. However, the large number of galaxies peaking over a wide range of redshift makes it difficult to employ traditional methods such as subtracting from the emission line flux the image of the stellar continuum \citep{2015A&A...580A..47V}. Furthermore, line fluxes will contribute to several J-PAS filters which also vary with the redshift of the object. Consequently, it is necessary to develop new techniques and algorithms in order to leverage completely the capability of J-PAS.
\\\\Machine learning techniques have effectively become a powerful tool over many fields where large quantities of data are available. The capability of these algorithms to find patterns in the data without making any empirical or theoretical assumption turns out to be their main advantage. In the last decades, astrophysical surveys are increasingly releasing vast amounts of data, which brings the opportunity of employing the most sophisticated up-to-date algorithms in order to analyse them faster and more efficiently. The applications range from the estimation of photometric redshifts \citep{2019A&A...621A..26P,2017MNRAS.465.1959C}, identification of stars \citep{2019A&A...622A.182W}, classification of galaxies \citep{2018MNRAS.476.3661D}, separation between galaxies and stars \citep{2020arXiv200707622B} to the determination of the SFR \citep{2019MNRAS.486.1377D,2019A&A...622A.137B} to cite some of the most recent research. In this work, we developed a new method based on Artificial Neural Networks (ANN) to detect and measure some of the main emission lines in the optical range of the spectrum: \Ha, \Hb, \nii, and \oiii. 
\\\\ This paper is organized as follows. We present in Sect. \ref{sec:surveys} J-PAS data together with data from other surveys that have been used to train and test the ANNs. In Sect. \ref{method} we describe in detail the main characteristics of the ANNs, how they can be trained and tested to deal with the uncertainties associated to the data. In Sect. \ref{sec:Simulations} we show the performance of ANNs in SDSS simulated data sets and discuss its main weakness. In Sect. \ref{aegisSDSS} we test our method in galaxies observed both in miniJPAS and SDSS. Finally, we summarize in Sect. \ref{sec:conclusion} and point out the steps needed to improve and extend the performance of the ANN in detecting and measuring emission lines.
\section{J-PAS and spectroscopic data}\label{sec:surveys}
In this section we present J-PAS and the spectroscopic data used throughout this paper for training and testing the model.
\subsection{J-PAS}\label{subsec:JPAS}
\begin{figure}
 \centering
 \includegraphics[width=\hsize]{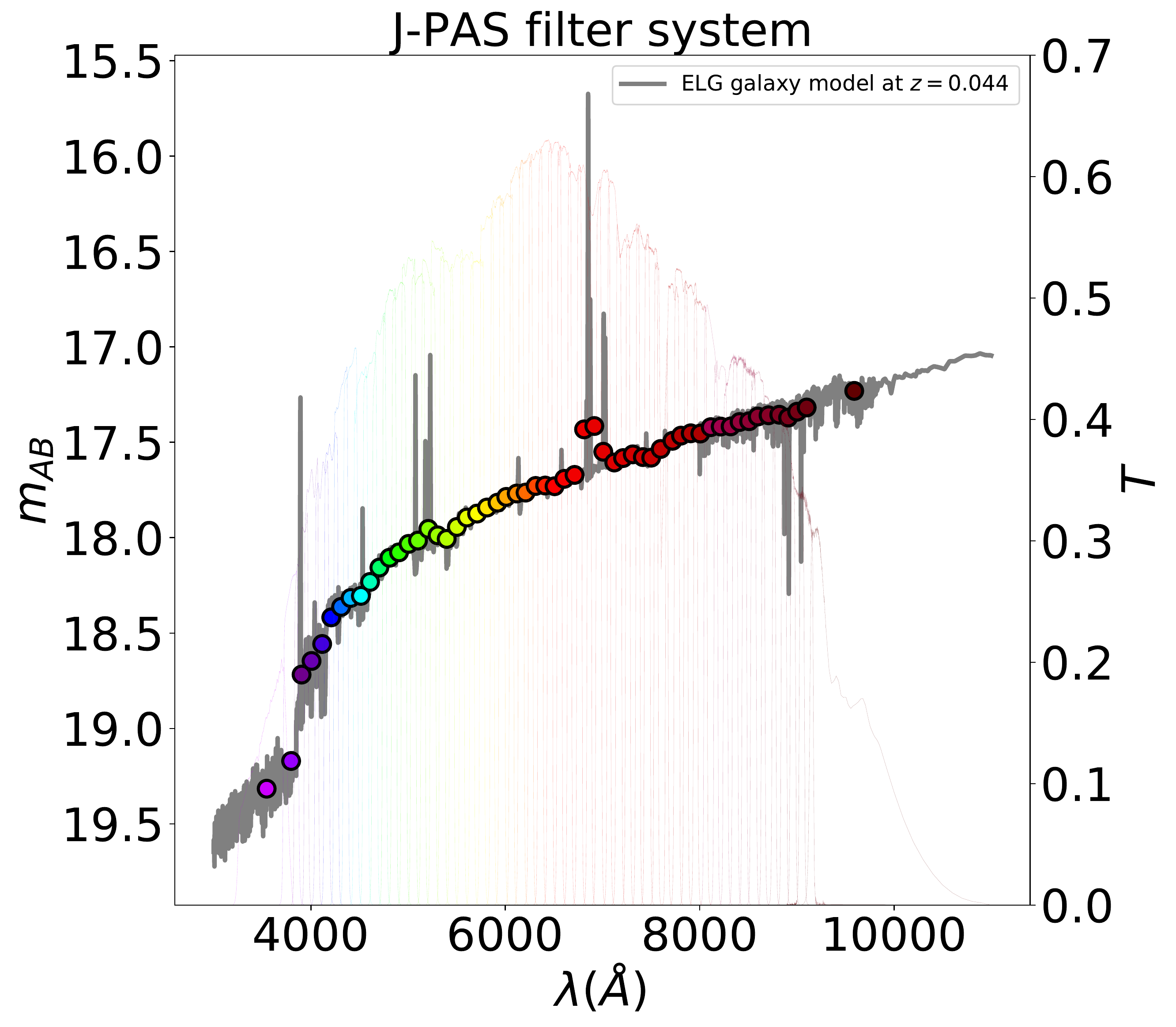}
  \caption{\tiny{Synthetic photometry (colored dots) of an emission line galaxy model (gray line) at $z=0.044$ in the J-PAS photometric system.}}
 \label{JPASexample}
 \end{figure}

J-PAS is an astrophysical survey  \citep{2014arXiv1403.5237B} planning to map $\sim8000$ deg$^2$ of the northern sky with 56 bands. This is, 54 narrow-band filters in the optical range plus 2 mediumband, one in the near-UV (uJAVA band) and another in the NIR (J1007 band). With a separation of $100$  \AA,  each narrow-band filter have a FWHM of $ \sim 145 $ \AA, whereas the FWHM of the uJAVA band is 495 \AA $ $, and the J1007 is a high-pass filter. The observations will be carried out with the $2.55$ m telescope (T250) at the Observatorio Astrofísico de Javalambre, a facility developed and operated by CEFCA,  in Teruel (Spain) using the JPCam, a wide-field 14 CCD-mosaic camera with a pixel scale of $0.2267$ arcsec and an effective field of view of $\sim4.7$ deg$^2$ \citep[see][]{2019A&A...622A.176C,2014JAI.....350010T,2015IAUGA..2257381M}. The survey is expected to detect objects with an apparent magnitude equivalent to $i_{AB} < 22.5$, up to $z\sim1$ with a photo-z precision of $\delta z \le 0.003(1+z)$ for luminous red galaxies. \\\\ The J-PAS project started its observations taking data with the Pathfinder camera observing four  AEGIS fields with 60 optical bands amounting to $1$deg$^2$. These data allow us to build a complete sample of galaxies up to $r_{SDSS} \le 22.5$ mag \citep{2020arXiv200701910B}. More than $60.000$ objects have been detected and can be downloaded from the website of the survey\footnote{http://www.j-spas.org/}. We describe more deeply the survey, referred as to miniJPAS, in Sect. \ref{subsect:miniJPAS} \\\\ One  example of how a nearby star-forming galaxy looks at the J-PAS resolution is shown in Fig. \ref{JPASexample}. The transmission curves of the J-PAS system are also shown in Fig. \ref{JPASexample}.
\subsection{CALIFA survey}
The Calar Alto Legacy Integral Field Area \citep[CALIFA,][]{2012A&A...538A...8S,2015A&A...576A.135G} is an integral field spectroscopy survey which observed 600 spatially resolved galaxies in the local universe ($0.005 <  z < 0.03$). The observations were taken with the $3.5$ m telescope at the Calar Alto observatory with the Postdam Multi Aperture Spectrograph \citep[PMAS,][]{2005PASP..117..620R} in the PPaK mode \citep{2006PASP..118..129K} which contains 331 fibers of $2.7"$ in diameter. With a field of view of $71'' \times 64''$ and a spatial sampling of $1$ arcsec/spaxel, CALIFA observed each galaxy in the wavelength range of $3700-7300 $ \AA $ $ with two different overlapping setups. Here we use the spectra taken in the low resolution setup (V500) that provides spectra from $3745$ to $7500 $ \AA $ $ with a spectral resolution of $6 $ \AA $ $ to generate J-PAS synthetic photometry. 
\\\\There are measurements of the emission lines available for a total of $275787$ spectra corresponding to 466 galaxies processed through the reduction pipeline of \cite{2015A&A...576A.135G}. These spectra include emission patterns of many different zones within the galaxy. Therefore, even though the integrated spectra of CALIFA galaxies might not be heterogeneous enough to build a training set, the individual zones cover plenty of diverse physical states. The properties of the stellar populations and the state of the ionized interstellar gas change from one region to another in each individual galaxy. Hence, with the amount of galaxies observed with CALIFA one can expect to have a rich representation of the most likely physical scenarios. The emission lines in each zone were measured from the residuals spectra obtained after subtracting the stellar continuum with {\sc Starlight} \citep{2005MNRAS.358..363C}. 

\subsection{MaNGA survey}\label{subsec:MANGA}
The Mapping Nearby Galaxies at Apache Point Observatory \citep[MaNGA,][]{2015IAUS..311..100B} is an ongoing integral field spectroscopic survey planing to observe spatially resolved spectra for ten thousand galaxies in the nearby universe ($z < 0.15$). With a wavelength coverage of $3600-10300$  \AA $ $ at a resolution of $R \sim 2000$, MaNGA is equipped with an IFU, in total $19$ fibers of $12 ''$ and $127$ of $32 ''$. In this work, we use the catalog available in \footnote{\url{https://www.sdss.org/dr14/manga/manga-data/manga-pipe3d-value-added-catalog/}} and processed by {\sc Pipe3D} pipeline in MaNGA SDSS-IV datacubes \cite{2016RMxAA..52..171S,2016RMxAA..52...21S}. The analysis of the stellar populations and ionized gas provides spatially-resolved information of the strongest emission lines in the optical range for a total of $4670507$ spaxels from 2755 galaxies.
\subsection{SDSS survey}
The Sloan Digital Sky Survey \citep[SDSS,][]{2000AJ....120.1579Y} contains spectroscopic measurements for more than three million astronomical objects and deep images of one third of the sky in five optical bands. The spectra were taken with a fiber of $3 ''$ in diameter and a spectral coverage of $3800-9200 $ \AA $ $ at a resolution of $R \sim 2000$. We use here the publicly available MPA-JHU DR8 catalog, from the Max Planck Institute for Astrophysics and the Johns Hopkins University \citep{2003MNRAS.341...33K,2004MNRAS.351.1151B}.  All the information regarding the catalog and the fitting procedure of the galaxy physical properties can be consulted online \footnote{\url{www.sdss3.org/dr10/spectro/galaxy_mpajhu.php}}. The catalog provides a total of $818333$ galaxies with redshift up to $z \sim 0.35$. We take only galaxies with reliable emission line measurements. As described in the data-model of the catalog, we can do that by excluding from the sample objects with \texttt{RELIABLE} $=0$ and/or \texttt{ZWARNING}  $>0$. We also discard galaxies where J-PAS synthetic magnitudes can not be calculated due to the lack of data in certain wavelength range of SDSS spectra. Finally, we end up with $701975$ galaxies.

\section{Method of analysis.}\label{method}
In this section we describe the architecture of the network in Sect. \ref{archi} and the strategies used for training and testing the model in Sect. \ref{trtest}. We also explain how to deal with photo-redshift uncertainty in Sect. \ref{subsec:Photo-redshift}, how errors can be estimated in Sect. \ref{errors}, and how to treat missing data in Sect. \ref{miss}. 
\subsection{Architecture of the Network}\label{archi}
\begin{figure}
 \centering
 \includegraphics[width=\hsize]{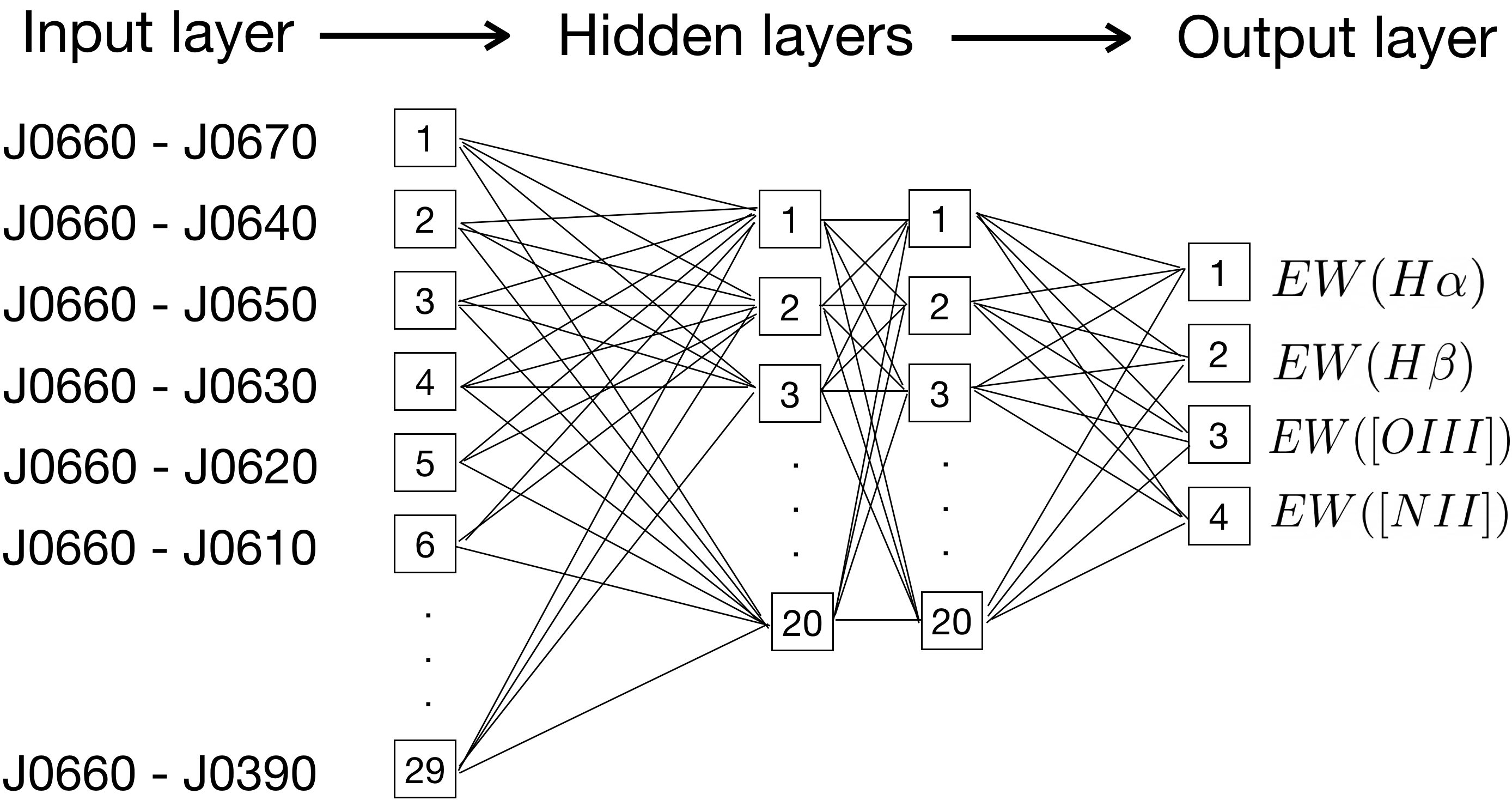}
  \caption{\tiny{Schematic diagram of the ANN$_R$ used for predicting lines emission at rest frame. The J0660 filter is our reference band for colors.}}
 \label{figNN}
 \end{figure}

In this paper we use a class of ANN called fully connected neural network. The implementation has been made with {\tt Tensorflow} \citep{tensorflow2015-whitepaper} and {\tt Keras} libraries \citep{chollet2015keras} in {\tt Python}. It is composed of a set of layers which have a specific number of neurons. The first layer contains the inputs (\textit{features}) of the network. In our application, the inputs are the colors of J-PAS measured with respect to the filter corresponding to \Ha for each spectrum. For instance, in nearby galaxies (z < 0.015) \Ha emission line will be captured by the J0660 band. Then, the color in the filter $Ji$ is defined as the difference respect to the magnitude measured in the J0660 band ($C_i = m_{AB}(J0660) - m_{AB}(Ji)$). The final layer contains the output of the network, sometimes also named \textit{targets} in the machine learning argot. Our targets are the equivalent width (EW) of \Ha, \Hb, \nii and \oiii. We built two different ANNs: one performs a regression task and obtains the values of these EWs, this network will be referred to as ANN$_R$. The other, ANN$_C$, carries out a classification between galaxies without emission lines (below a given threshold) and emission line galaxies by imposing cuts in the EWs of the mentioned lines. We could have performed this classification based on the values yielded by the ANN$_R$ but an algorithm specifically constructed for that will always obtain better results.
\\\\ As we mentioned before, emission line fluxes have contribution to different bands according to the redshift of the source and the width of the emission line. The redshift might be treated as an input in the model but that would imply to train the ANN with a uniform distribution in this parameter, otherwise the ANN would not be able to predict equally at all redshifts. Furthermore, this approach would reduce our sample size and limit our range of predictability due to the different redshift coverage of CALIFA, MaNGA and SDSS. For these reasons, we train a different ANN for each redshift, going from $0$ to $0.35$ with a step of $0.001$. We shift all the spectra of the training set in wavelength at the same redshift and we compute the colors within the common wavelength range between J-PAS and the spectroscopic surveys described in Sect \ref{sec:surveys}. This range depends on the redshift and consequently the number of inputs vary between 28 and 39 colors.
\\\\Between the input and the output layers the ANN can hold inner layers, commonly called \textit{hidden} layers, with absolute freedom to decide the number of layers and neurons in it. There is no standard recipe to find the optimal architecture of a network. Theoretically, with one hidden layer, it is possible to model the most complex function with sufficient amount of neurons. However, deep ANN with mores hidden layers have a much higher parameter efficiency and can hence model complex functions by using much less neurons \citep{geron2019hands}. Therefore, few hidden layers are normally sufficient if the relation between inputs and output is not very complex. Certainly, this is our case because the emission lines are clearly visible in the J-spectra. Besides, other features such as the color of the spectra, that can help as well to estimate the emission line patterns, are linearly connected to the inputs. In other cases where the relation is more complex, for instance the estimation of the photometric redshift based on the images of a galaxy \citep{2019A&A...621A..26P}, an architecture of many more hidden layers will converge faster obtaining better performance. 
\\\\ The amount of neurons in the hidden layer varies between the size of the input and the size of the output layers. Our ANNs have 2 hidden layers with 20 neurons each, which is in between the number of inputs (34 colors in average) and the number of outputs (four EWs for the $ANN_R$ and two classes in the case of the $ANN_C$). A schematic view of the ANN$_R$ used in this work can be seen in Fig. \ref{figNN}.
\\\\All the neurons in a given layer are connected to the neurons in the contiguous layer by a matrix of weights \textbf{W} and a bias \textbf{B}:
\begin{equation}
\mathbf{L_{n}} = g(\mathbf{W_n}  \cdot  \mathbf{L_{n-1}} + \mathbf{B_n})
\end{equation}
where $\mathbf{L_n}$ refers to layer $n$. $L_{0}$ are the inputs of the ANN and $g$ is the activation function of neurons. It worth mentioning the importance of such function, being responsible for the non-linear behavior in the network. Otherwise, the outputs would be simply a linear combination of the inputs, which would not be sufficient to address most of the problems. We use the so-called Rectified Linear Unit (ReLU) activation function \citep{Relubibcode}, which has become the default activation function in recent years due to its advantages \citep{pmlr-v15-glorot11a}.
\\\\ Typically, ANN are trained using an algorithm commonly referred to as \textit{backpropagation}.  Adjusting the set of weights and bias that minimizes a certain \textit{loss-function} is the actual process of training. For regression-like problems the most common loss-function is usually a \textit{mean square error}, while for binomial classification the \textit{binary cross entropy} is frequently employed. We make use of these functions in our models.
\\\\ One important aspect to take heed of when when we are training an ANN is to avoid  \textit{overfitting}. Improving the loss-function indefinitely will make the algorithm to fit features of the data that do not represent the general trend. Consequently, the predictability of the network will be compromised. We can avoid that by imposing a maximum value over the weights that each neuron can carry. \\\\
Optimising the architecture of the network is a process that requires tweaking many hyper-parameters. Along this work we have tested different architectures, increasing and decreasing the number of neurons and/or hidden layers or by using alternative loss functions such as the \textit{mean absolute error} or the \textit{mean relative error} for regression. Sometimes even different architectures can obtain very similar results. The model that we describe in this paper is among the ones we tested that better perform.
\subsection{Training strategy}\label{trtest}
 We generate synthetic J-PAS data by convolving the spectra presented in Sect. \ref{sec:surveys} with the J-PAS filter system. Since CALIFA, MaNGA, SSDS, and J-PAS have different wavelength coverage, we only use in our model the common wavelength range of the four instruments at $z = 0$, which is $3810 - 6850$  \AA.
\\\\ The training sample is built differently depending on whether we are dealing with a classification or a regression task. In a classification problem an unbalance number of classes in the training sample might under-predict the minor class \citep[see e.g.][for a review in the topic]{Ali2015ClassificationWC}. Therefore, in cases where it is possible, a balance training set is more desirable. In regression-like problems the optimal training set is the one that better cover the parameter space of the target variables. For instance, a training set built for classifying galaxies above and below 3 \AA $ $ in the EW of \Ha will be different from one that aims to compute the same EW in the range between 0 and 20 \AA. Simply because one would need much more galaxies in the interval from 3 to 20 \AA $ $ than the ones needed below 3 \AA.
\\\\ Considering the data that we have at hand, there are other aspects that need to be taken into account to build the training sample. Firstly, in order to ensure the algorithm receives the most reliable information, one would desire to select only the spectra where emission lines have been measured with high signal-to-noise ratio (SN). However, being too strict in the selection criterium induces a bias towards  line-emmiting galaxies and reduces significantly the size of the sample. Secondly, while CALIFA and MaNGA have observed the nearby universe resolving spatially the physical properties of the interstellar medium within galaxies, SDSS can only see the inner parts of nearby galaxies but with the advantage of covering distances further away in the universe. It has been shown how spatial resolution affects the location of points (spaxels) in the BPT, possibly altering AGN classification and/or simulating it via mixed spectral featured \citep{2016A&A...586A..22G}. Finally, the emission lines catalogs obtained from these surveys have been derived with different fitting tools which makes it difficult to compare them in equal terms. 
\\\\In essence, there is not a simple and unique way of putting together all these data and build the training set that better represents the universe as J-PAS will look at it. Instead, we propose to train the ANN with different training sets in order to understand the source of errors and inaccuracies of the model.
\subsubsection{Training and testing sets in the ANN for classification}\label{trcl}
With the aim of identifying galaxies with low and high emission lines, we train a ANN classifier to perform a binary classification based on the EW of \Ha, \Hb, \nii or \oiii. This type of classification might allows us to disentangle the structure of the bimodal distribution found in the EW of \Ha in CALIFA and SDSS galaxies \citep{10.1111/j.1365-2966.2008.13963.x,2018MNRAS.474.3727L}. In these works the authors found that the mentioned bimodal distribution has its minimum around 3 \AA $ $. In the regime of low emission the J-PAS filter system is not sensitive enough to detect emission lines and hence, only via machine learning, which can extract features from the J-spectra much more complex, it is possible to address this problem.
\\\\Galaxies are considered emitting-line galaxies or $Class\, 1$ according to the following criteria:
\begin{equation}
\begin{split}
EW(\Ha) > EW_{min} \parallel EW(\Hb) > EW_{min} \parallel \\ EW(\oiii) > EW_{min} 
\parallel EW(\oiii)> EW_{min} 
\end{split}
\end{equation}
and $Class$ $2$ in the rest of cases. We train several classifiers where $EW_{min}$ takes the following values: 3, 5, 8, 11, and 14  \AA. In short, if a galaxy has an EW greater than the $EW_{min}$ in any of these lines, it will be considered as $Class\, 1$. If all the EWs in a galaxy are below the threshold then it will be tagged as $Class\, 2$. 
\\\\In most of the cases \Ha is the most powerful emission line and consequently it will decide whether galaxies belong to one class or other. There is nothing special in the values chosen for $EW_{min}$ except that they are in the regime of low emission. With the ANN classifier we proof that this regime can be explored in J-PAS and any other $EW_{min}$ around these values could be implemented in the future.
\\\\ The combination of data from different surveys used in this work does not improve or worsen the performance of the ANN classifier. Consequently, for the sake of simplicity, we train only with CALIFA synthetic J-spectra and we test with SDSS galaxies. We do not impose any cut in the errors of the EWs, but we ensure to have the same amount of J-spectra in both classes in the training set. We end up with $200000$ synthetic J-spectra to perform the training. 
\subsubsection{Training and testing sets in the ANN for regression}\label{trreg}
For the purpose of obtaining the values of the EWs of galaxies in J-PAS, we propose two training sets. The first one, what we call the CALMa set, is only composed of CALIFA and MaNGA synthetic J-spectra while the second one, the SDSS set, includes only SDSS galaxies.
\\\\We test the performance of the model by removing randomly 15000 synthetic J-spectra from the training samples: 5000 from CALIFA, 5000 from MaNGA and 5000 from SDSS. Those synthetic J-spectra are considered as validation or test samples depending on the training sample. For instance, if we train with the CALMa set, we use MaNGA and CALIFA samples to tune the hyper-parameters of the model (validation samples) and SDSS galaxies to actually evaluate the model; and the other way around: if we train with the SDSS sample, SDSS galaxies plays the role of the validation sample and CALIFA and MaNGA synthetic J-spectra are used for testing purpose. In this way, we ensure that the color terms that might appear as a result of fitting tools used to derive the emission lines and/or the instruments that took the spectra are not playing a mayor role in the prediction made by the ANN. If that were the case, building samples with different surveys in the training and testing sets allows us to identify any potential bias of such origin.
\\\\ We add to the training set only those synthetic J-spectra where emission lines have an error below a certain threshold. In the case of MaNGA galaxies, spaxels with signal-to-noise-ratio (SN) below $10$ in the flux of \Ha, \Hb, \nii or \oiii are discarded. However, we were more flexible with spaxels in CALIFA and SDSS galaxies, going down to a SN of $2.5$. Such flexibility allows us to increase the amount of low-emitting galaxies in the samples. In addition, when it comes to the CALMa set, we achieve a more equilibrated weight between the prominence of CALIFA and MaNGA in the training sample. We also exclude from the training set the spectra where the EWs are greater than $600$ \AA (these are very rare cases, 10 in total). Since the loss function is quadratic in the EWs, such type of spectra force the ANN$_R$ to fit at the same time two antagonistic regimes: low-emitting and extreme emission line galaxies. Consequently, it would worsen the performance of the ANN$_R$ in the range of interest. Finally, we end up with a training set of $134000$ synthetic J-spectra from CALIFA, $280270$ from MaNGA, which together form the CALMa set; and $135300$ galaxies in SDSS set.

\subsection{Photo-redshift uncertainty}\label{subsec:Photo-redshift}
Even though J-PAS will provide redshifts with high precision \citep[$\delta z \le 0.3 \% $ \footnote{Throughout this paper we use the convention $\Delta z = (1 + z)\delta z$, where $\Delta z = z - z_{photo}$.} for luminous red galaxies]{2014arXiv1403.5237B}, the performance of the ANN could be compromised in many cases. 
Let us assume for example that we aim to compute the EWs of a galaxy at redshift $0.3$ with $\Delta z = 0.003$. In the best case scenario, the galaxy redshift would be between 0.296 and 0.304. According to our redshift bin, we have 8 possible ANNs to try with. While in the vicinity of the true redshift the ANN can reasonably make a good job, in the extremes the EWs would dramatically be underestimated.  Since colors are computed with respect to a filter far away from the one corresponding to \Ha, the ANN will interpret as an absorption line what indeed is an emission line. Although the probability density functions (PDFs) of the photo-z can help improving the predictability assigning weights to each redshift, whenever we found a non-gaussian PDF with, for instance, an asymmetric distributions with two peaks, it would be difficult for the ANN to make reasonable predictions. 
\\\\A way to obtain better results in galaxies where the uncertainty in the redshift if high is to consider only the configurations (redshifts) that maximizes a certain function. Certainly, for emission line galaxies, the redshift where the sum of all EWs reaches the highest value is close to the true redshift. However, this redshift overestimates the EWs in galaxies with low emission. In order to minimize such effect, we average over the five configurations (redshifts) that maximize the sum of all EWs within the photo-redshift uncertainty ($\Delta z$). The fact that these configurations might be found in non-contiguous redshift bins can help in those cases where there are asymmetric PDF distributions of photo-redhshifts. 
\\\\As we discuss latter in Sec. \ref{subsec:test5max}, this method is able to somehow recompute the distance of the galaxy correcting a possible deviation from the spectroscopic redshift in galaxies where $\sum EW_i > 20 $ \AA. Therefore the method of the five maximum, hereafter \textit{5max}, can certainly help the ANN$_R$ to improve its performance but cannot be used with the ANN$_C$. Most probably, it would increase the amount of false positives as the redshift uncertainty increases. In Sect. \ref{sec:Simulations} we quantify how the error in the redshift can impact the predictions of the ANN$_C$ and the ANN$_R$. Fortunately, the ANN$_C$ is less sensitive to that (see Fig. \ref{figROC} and Table \ref{areasroc}).
\subsection{Estimation of errors}\label{errors}
The uncertainty of the ANN method can be estimated considering three sources of errors: the error of the photometry, the error in the photometric redshift, and the intrinsic error of the ANN training.  Before the training actually starts, weights and biases in ANN can be set to a certain value by initialising randomly according to any distribution function. Generally, each initialization state will converge to different local minimum of the loss-function. Even though it is possible to find the state that leads to the best score over the validation sample, usually a Monte Carlo approach called the \textit{committee}, this is, the mean of the individual predictions of a set of ANN, will be a more robust and accurate estimate of the targets. Then, the variations of the outputs in each individual member of the committee respect to the mean provide an estimation of the uncertainty in the predictions intrinsically associated to the training procedure. The list bellow details the steps to follow in order to account for the contribution of each uncertainty to the errors budget. \\
\begin{enumerate}
	\item Photometric error: we input the ANN with $N+1$ different values of the magnitude, where one corresponds to the nominal value and the other N are randomly drawn from a gaussian distribution centred on the nominal value and with standard deviation equal to the photometric error. The median (M) and the  median absolute deviation (MAD) of N+1 predictions give us the prediction and the weight of one member in one committe:\\
	\begin{equation*}
	    \begin{split}
            P_{iz_j} & = \text{M} [p^{iz_j}_{0},p^{iz_j}_{1},...,p^{iz_j}_{N+1}] \\
            W_{iz_j} & = 1/\text{MAD}[p^{iz_j}_{0},p^{iz_j}_{1},...,p^{iz_j}_{N+1}]
        \end{split}
	\end{equation*}
	where i stands for the committe member and $z_j$ for the redshift.\\
	\item ANN intrinsic error: the prediction of the committe in a given redshift can be estimated by computing the average (AVG) of all members in the committe with the weights obtained above. The error of the committe is simply the MAD of m(N+1) prediction, where m refers to the number of members in the committe. We found that averaging over five members is enough to obtain reliable results: \\
		\begin{equation*}
	        \begin{split}
	            P_{z_j}  & = AVG[P_{0z_j},P_{1z_j},...,P_{mz_j};W_{0z_j},W_{1z_j},...,W_{mz_j}] \\
	            \epsilon^{ANN}_{z_j}  & = \text{MAD} [p^{1z_j}_{0},...,p^{1z_j}_{N+1},p^{2z_j}_{0},...,p^{2z_j}_{N+1},...,p^{mz_j}_{0},...,p^{mz_j}_{N+1}] \\
            \end{split}
	    \end{equation*}
	\item Photo-redshift uncertainty: we compute the median value of $n$ committes, one for each redshift. In the case of the ANN$_R$ we select the five maximum setting (see Sect. \ref{subsec:Photo-redshift}) and for the ANN$_C$ we consider all the redshift within the error range.\\
	    \begin{equation*}
	        \begin{split}
                P_{ANN_R} & = \text{M} [P_{z_0}(max_0),P_{z_1}(max_1),...,P_{z_4}(max_4)] \\
                P_{ANN_C} & = \text{M} [P_{z_0},P_{z_1},...,P_{z_n})]
            \end{split}
	    \end{equation*}
	Finally, the error is the quadratic sum of the median error of all committees plus the dispersion of these committees respect to the median, which gives us the contribution of the redshifts uncertainty. \\
	    \begin{equation*}
	        \begin{split}
                \epsilon_{ANN}  & = \sqrt{\text{M}[\epsilon^{ANN}_{z_0},\epsilon^{ANN}_{z_1},...,\epsilon^{ANN}_{z_n}]^2 + \text{MAD}[P_{z_0},P_{z_1},...,P_{z_n}]^2}\\
              \end{split}
	    \end{equation*}
    If the spectroscopic redshift of the object were known, the expression above would be simply: $\epsilon_{ANN} = \epsilon^{ANN}_{z_{spec}}$
\end{enumerate}

\subsection{Missing data}\label{miss}
Many are the problems, both related to the data reduction or the observation, that could lead to incomplete or missing data. Consequently, a fraction of our sample will lack photometric measurements in some of the filters used by the ANN. Certainly, many of such objects will have to be rejected automatically if the photometry is not reliable in the bands capturing the emission lines. However, there will be galaxies where the photometry might be problematic only in the some of the bands dominated by the stellar continuum. For instance, in the miniJPAS area, among the galaxies that are below 0.35 in redshift and 22.7 magnitudes in the rSDSS band (2291), $30 \%$ of them have at least one band where the photometry is not reliable. Most of the galaxies in this sample (70 $\%$) have a median SN ratio below 10. Naturally, this fraction will decrease as the median SN of the sample increases. 
\\\\One solution  to address the problem of missing date requires training several ANN considering different configurations where part of the data is accessible. Nevertheless, this would imply testing the performance of the ANN in many scenarios and would be computationally very expensive. The other solution is to replace the missing data in the corresponding filter with the fluxes obtained from the spectral fitting of the stellar continuum. Several spectral fitting codes can be used such as \texttt{MUFFIT} \citep{2015A&A...582A..14D} or \texttt{BaySeAGal} (Amorim et al. in prep.). This analysis provides reliable photometric predictions for the missing data, as well as information regarding their stellar population properties (e.g.~stellar mass, age, and extinction, which is always necessary for a more comprehensive picture). Furthermore, the stellar continuum is needed for obtaining absolute emission line fluxes. We follow this technique to treat the missing data in J-PAS.
 \section{Validation of the method.}\label{sec:Simulations}
In this section we perform several tests to study the predictability and limitations of the model. Firstly, we evaluate the capability of the ANN$_C$ in Sect. \ref{sec:class}. Secondly, in Sect.\ref{EWz}, we compare the predictions of the EWs obtained by the ANN$_R$ and trained with the CALMa set with the SDSS testing sample. Then, In Sect. \ref{subsec:table} we compare the performance of the different training sets proposed in Sect. \ref{trreg}. In Sec. \ref{subsec:test5max} we test the $5max$ method and we study in Sect. \ref{subsec:EWdependence} the impact of the redshift uncertainty on the ANN$_R$ predictions as a function of the EW. Finally, in Sec. \ref{subsec:EWlimit} we estimate the minimum EW measurable in function of the SN of the photometry for each of the emission lines predicted by the ANN. 
 \subsection{Classifying galaxies}\label{sec:class}
\begin{figure}
 \centering
 \includegraphics[width=\hsize]{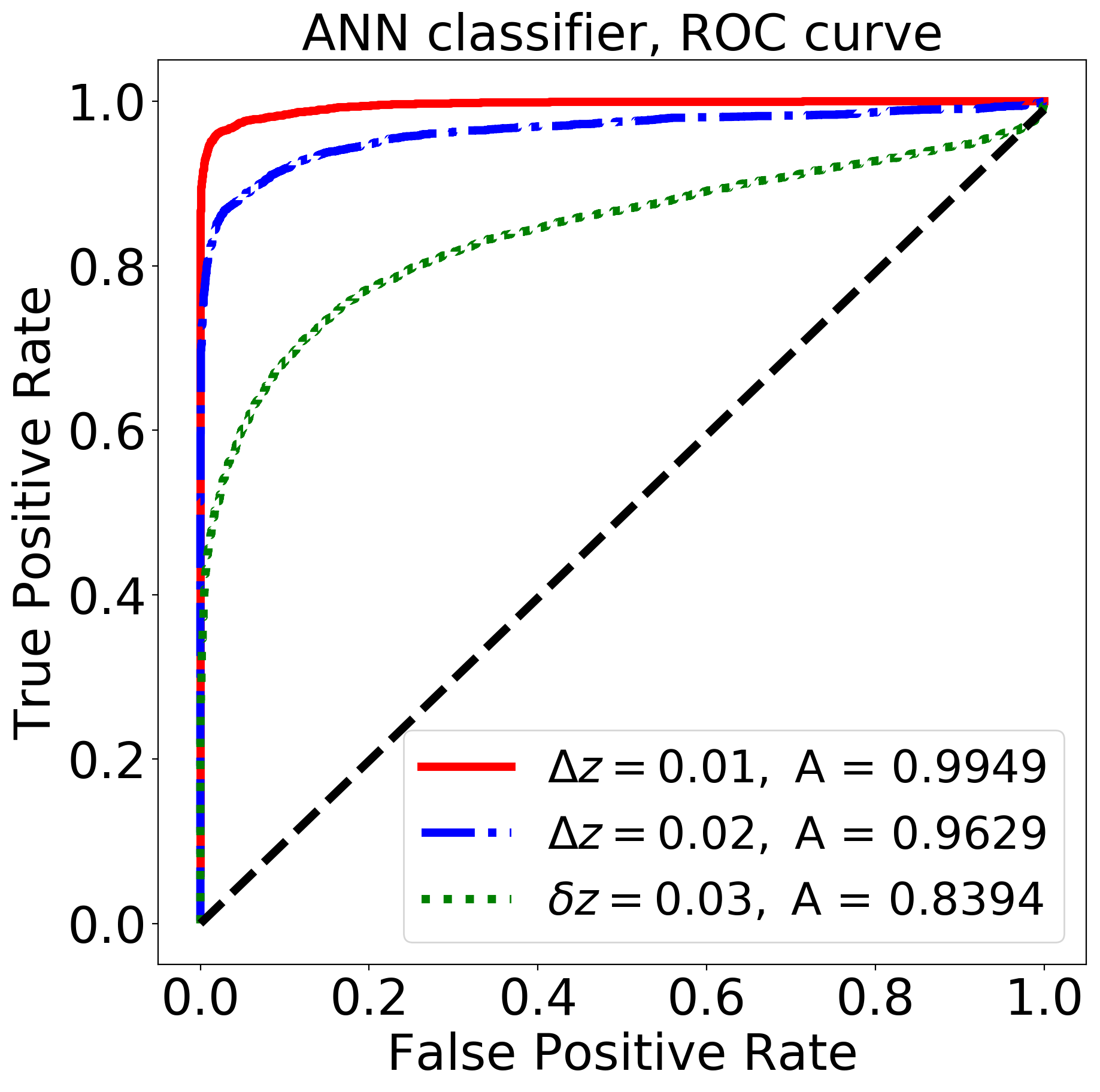}
  \caption{\tiny{ROC curve of the ANN$_C$ for $EW_{min} = 3$ \AA $ $ as a function of the redshift uncertainty for 10000 SDSS galaxies. The legend shows the areas under the ROC curves for each $\Delta z$. In Table \ref{areasroc} we show these values for other $EW_{min}$ settings. Blue dashed line shows the performance of a random classifier.}}
 \label{figROC}
 \end{figure}

 The ANN$_C$ is trained with the CALIFA training sample. For evaluating its efficiency, we
 explicitly selected a subset of 10000 galaxies from the SDSS catalog with the same amount of classes, this is: $5 000$ galaxies that belong to $Class 1$ and  $5 000$ to $Class 2.$ (see Sect. \ref{trcl}). Galaxies in each class are picked at random from the entire catalog. For each galaxy the ANN$_C$ yields a number between 0 and 1 indicating the probability of being one of the two classes.  As we discuss in Sect. \ref{subsec:test5max}, the \textit{5max} method (Sect. \ref{subsec:Photo-redshift}) is not suitable for galaxies without emission lines. Most probably, it would increase the amount of false positives as the redshift uncertainty increases. Since we have noticed that the ANN$_C$ is less sensitive to redshift and is able to classify galaxies even when its uncertainty is high, we simply compute the average of each one of the predictions within the redshift interval defined by $\delta z$.
 \\\\We show in Fig. \ref{figROC} the receiver operating characteristic (ROC) curve, which represents the true positive rate (TPR) versus the false positive rate (FPR) for $EW_{min} = 3 $ \AA. We also show how the ROC curve varies as a function of the redshift uncertainty. The ANN$_C$ scores very high even when $\delta z = 0.01$ and loses efficiency gradually as the uncertainty in the redshift increases. We summarize in Table \ref{areasroc} the area under the ROC curves for others $EW_{min} $. The ROC curves do not show remarkable changes in function of the $EW_{min}$ used in the classification.

\begin{table}
\tiny
\begin{tabular}{|l|l|l|l|}

\hline
$EW_{min}$ & Area ($\Delta z = 0.01$)&  Area ($\Delta z = 0.02$) &  Area ($\Delta z = 0.03$) \\ \hline
 3 \AA & 0.9949 & 0.9629 &  0.8394 \\ \hline
  5 \AA & 0.9948 & 0.9507 & 0.8160 \\ \hline
  8 \AA &  0.9938& 0.9604  & 0.8407 \\ \hline
 11 \AA  & 0.9915 & 0.9594 & 0.8547 \\ \hline
  14 \AA & 0.9894 & 0.9600 & 0.8614 \\ \hline
\end{tabular}
\caption{\small{Area under the ROC curve as a function of the redshift uncertainty 
and the $EW_{min}$ used in the classification.}}
\label{areasroc}
\end{table}

 \subsection{Emission-line galaxies: EWs, line ratios and BPT diagram}\label{EWz}
In this section we discuss how the CALMa training set (see Sect. \ref{trreg}) scores in the SDSS testing sample. We use the spectroscopic redshift provided in the catalog without considering any error so as to separate the uncertainties intrinsically associated to the model from those related to redshift. We do not consider the errors of SDSS spectra, we rather add gaussian noise to each magnitude 100 times assuming an average SN of 10. This allows us to treat all galaxies in the same manner and assume higher errors.
\\\\ The testing set from CALIFA, MaNGA, and SDSS are composed of $5 000$ synthetic J-spectra with SN in the EWs above $10$. This criterion excludes many galaxies with low-ionization nuclear emission-line region (LINER). We also exclude the spectra where the EWs are greater than 600 \AA $ $ to test the model in the range of which we trained the ANN$_R$. Hence, even though we are able to identify strong and weak emission lines galaxies, their EWs might not be accurate due to these selection criteria on the training sample.
\subsubsection{Equivalent widths}\label{EWsub}
\begin{figure*}
    \includegraphics[width=0.94\linewidth]{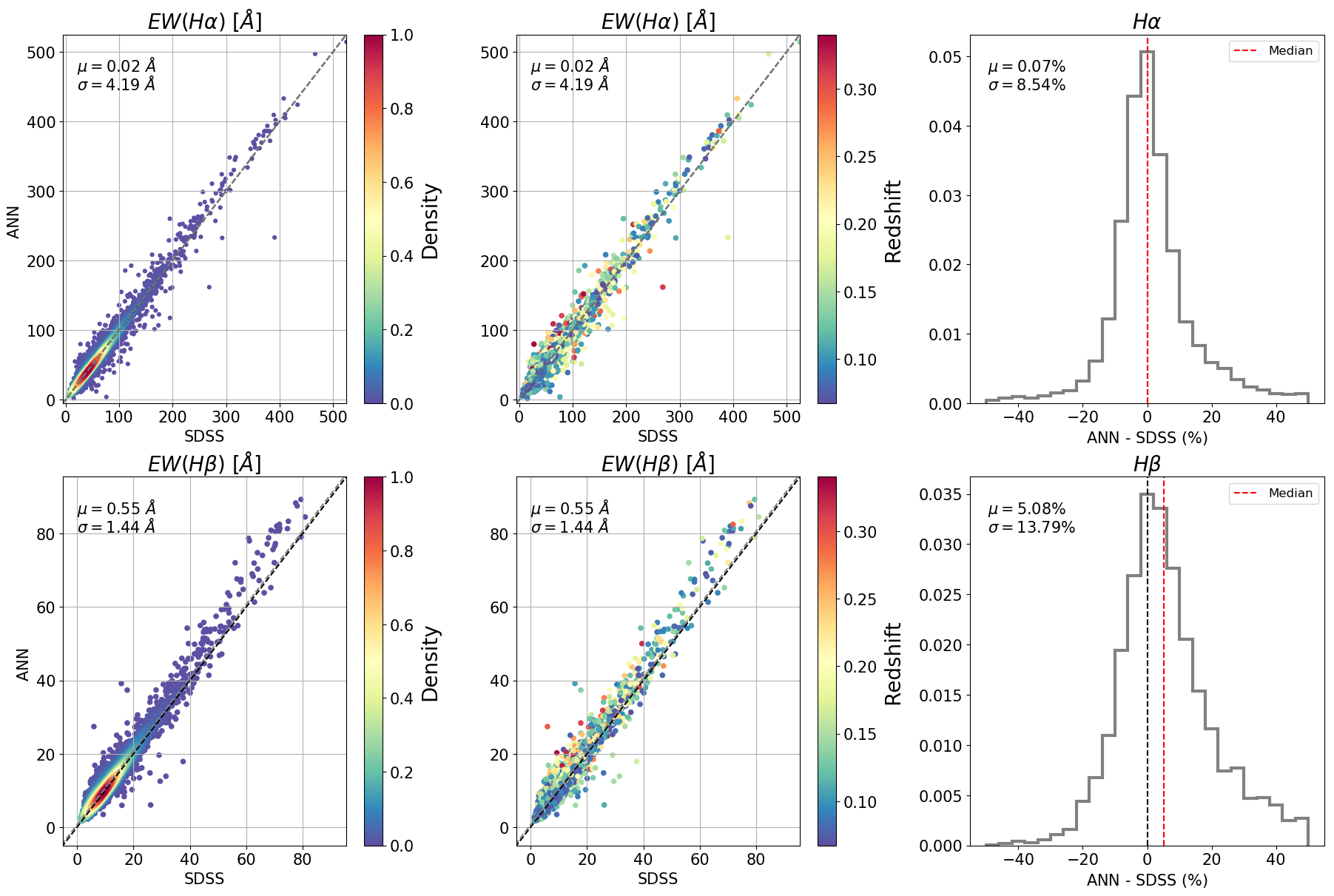}
    \includegraphics[width=0.94\textwidth]{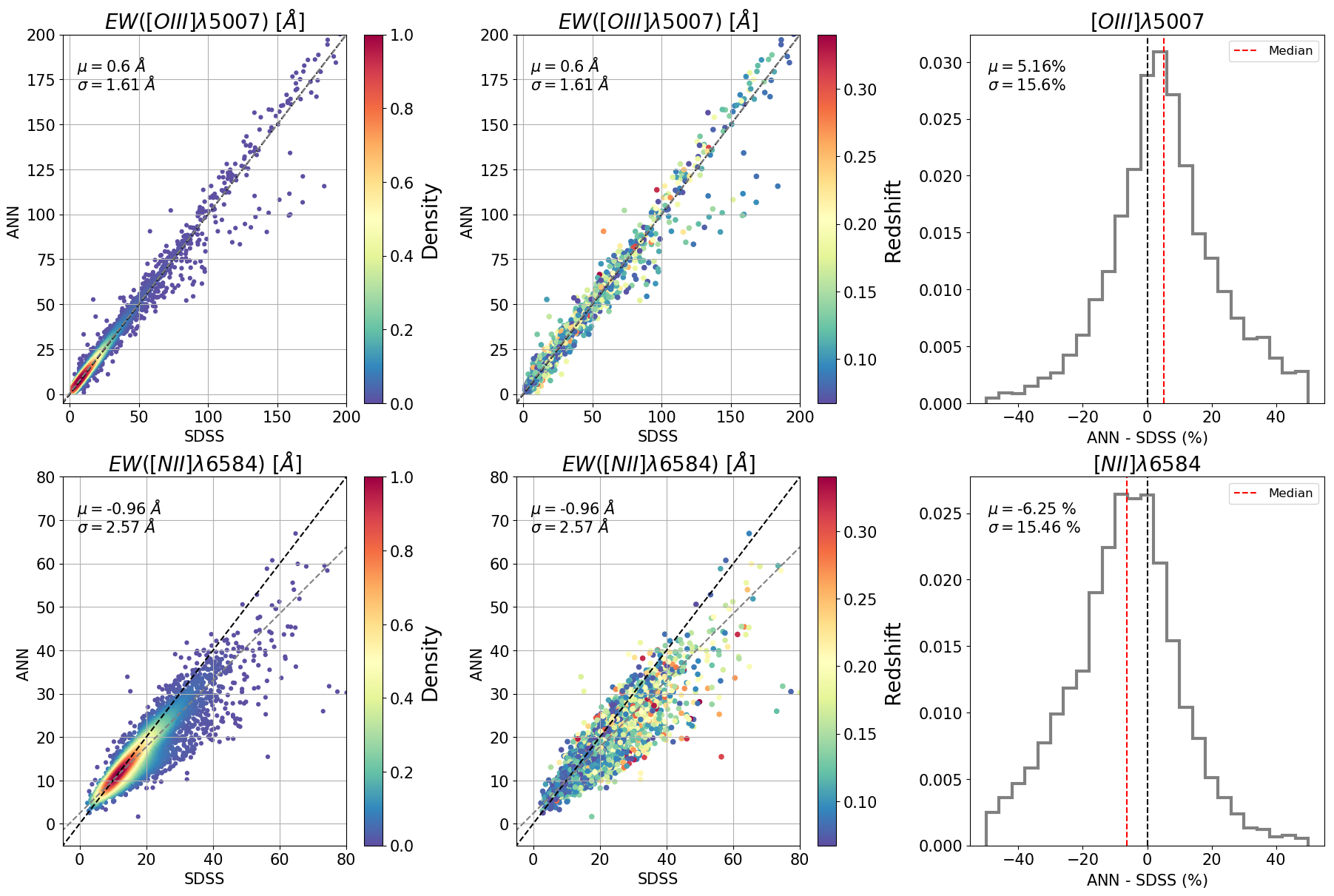}
 \caption{\tiny{EWs of \Ha, \Hb, \nii and \oiii predicted by the ANN$_R$ compared to SDSS testing sample. The ANN$_R$ is trained with the CALMa set. The color-code represents the probability density function defined by a Gaussian kernel  (right panel) and the redshift of the objects (left panel). The histograms in are normalized to one and show the relative difference between both values. Black and blue numbers are the median and the median absolute deviation of the difference. Black and grey dashed lines on the left are lines with slope one and the best linear fit respectively. We perform a sigma clipping fit with $\sigma = 3$ to exclude outliers. Red dashed line represents the median.}} 
\label{fig:lines}
\end{figure*}

 Fig. \ref{fig:lines} compares the EWs predicted by the ANN$_R$ and those in the SDSS testing sample (extracted from the MPA-JHU DR8 catalog). 
 We do not plot the errors yielded by the ANN$_R$ for visual reasons. A complete analysis of the errors estimated by the ANN$_R$, as discussed in Sec. \ref{errors}, is performed in Sec. \ref{subsec:EWlimit}. The plots on the left are color-coded with the density of points and the ones in the middle with the redshift of the galaxy. The histograms on the right represents the relative difference between the ANN$_R$ predictions and the SDSS testing set. We constrain better the EW of \Ha followed by \Hb, \oiii and \nii (see median and median absolute deviation in Fig. \ref{fig:lines}). The \Ha line, which is the most powerful one, presents less dispersion and bias. \Hb and \oiii lines are recovered with similar precision and \nii line shows more dispersion and bias. We observe that \nii line saturates at high values, that is to say, the EWs tend to be underestimated as the strength of the line increases. The same effect occurs in the \oiii line in form of a second branch. We analyze this effect in Sect. \ref{ratios}. We do not observe strong color gradients in the plots color-coded with the redshift, indicating we are not biased regarding the distance of the objects.
 \\\\ In summary, the EWs of \Ha, \Hb, \nii, and \oiii can be predicted with a relative standard deviation of $8.5 \%$, $13.8 \%$, $15.6 \%$, and $15.5 \%$ respectively. \Ha, \Hb, \nii, and \oiii lines presents a relative bias of $0.1\%$, $5.1 \%$, $5.2 \%$, and $-6.3 \%$ respectively. In a future work, we will study the distribution of all these values using a real and complete  sample of galaxies from miniJPAS.

\subsubsection{Ratios between emission lines}\label{ratios}
\begin{figure*}
    \includegraphics[width=\linewidth]{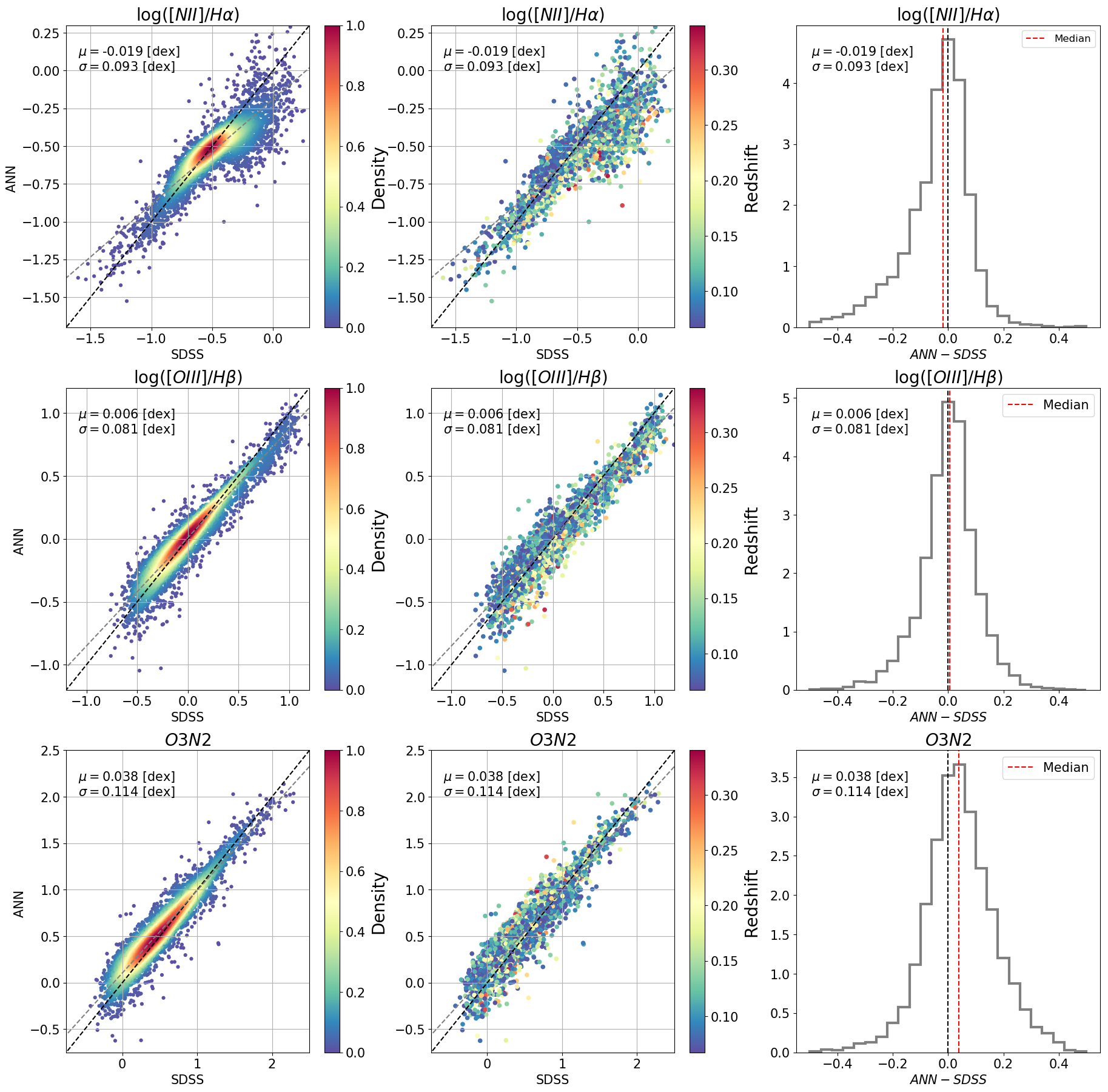}
 \caption{\tiny{Comparison between \nii/\Ha, \oiii/\Hb and $ \ion{O}{3}\ion{N}{2}$ ratios estimated by the ANN$_R$ and SDSS testing sample. Same scheme of Fig. \ref{fig:lines}. The ANN$_R$ is trained with the CALMa set.}}
\label{fig:ratios}
\end{figure*}

From the EWs we can easily obtain the ratios of \nii/\Ha and \oiii/\Hb under the approximation that each couple has the same stellar continuum. From that, we also obtain the metallicity indicator  \ion{O}{3}\ion{N}{2} $\equiv \log \{(\oiii/\Hb)/(\nii/\Ha)\}$ \citep{2004MNRAS.348L..59P}.
Fig. \ref{fig:ratios} shows the comparison between the logarithmic ratios obtained with ANN$_R$ and the SDSS testing sample. As in Fig. \ref{fig:lines} the plots are color-coded with the density of points (left column) and the redshift of the galaxy (middle panel). The histograms on the right show the logarithmic difference between the ANN$_R$ predictions and the SDSS testing set.
  \\\\ The \nii/\Ha ratio is predicted within $0.093$ dex and a bias of $-0.019$ dex. The \oiii/\Hb ratio is slightly better constrained, with no bias and a dispersion of $0.081$ dex. Finally, the $ \ion{O}{3}\ion{N}{2}$ is recovered within $0.114$ dex and a bias of $0.038$ dex. The saturation of the \nii line at high values is responsible of the same effect observed in the \nii/\Ha ratio. Since MaNGA and CALIFA surveys observed galaxies spatially resolved, the number of star-forming regions is much more numerous in the training sample and consequently the ANN$_R$ has few spectra to constrain the ratio of \nii/\Ha in galaxies hosting an AGN. To a lesser extent, that also occurs as well in the \oiii/\Hb ratio for galaxies with values higher than $3.2$ and in form of a second branch in the \oiii line.
 \subsubsection{BPT diagram}\label{BPTSect.}  
\begin{figure*}
    \includegraphics[width=\linewidth]{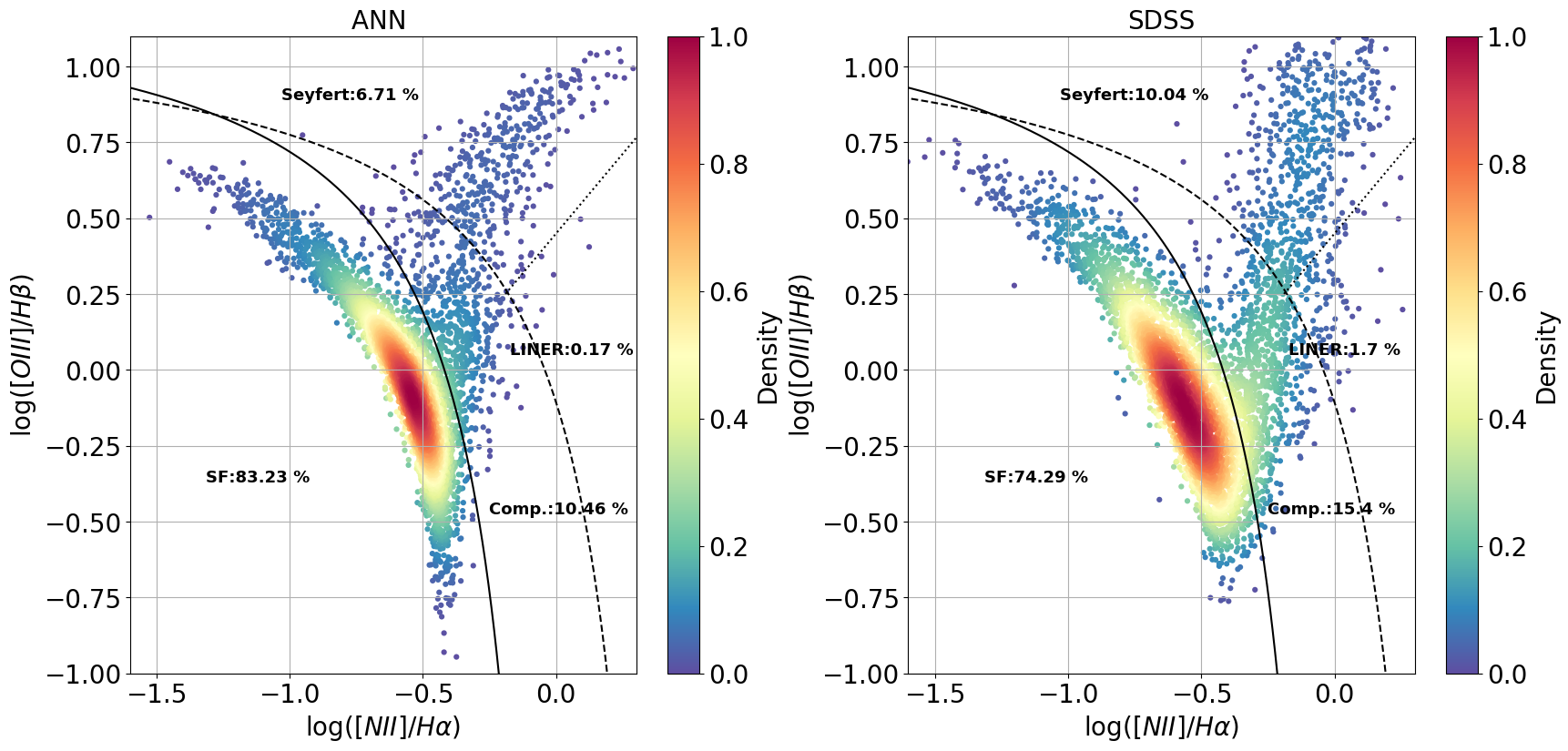}
 \caption{\tiny{BPT diagram obtained with the ANN$_R$ and SDSS testing sample from the MPA-JHU DR8 catalog. The ANN$_R$ is trained with the CALMa set. The color-code indicates the density of points. The solid (ka03), dashed (Ke01) and dotted lines (S07) define the regions for the four main ionization mechanism of galaxies. The percentage for each group is shown in black.} }
\label{fig:BPT}
\end{figure*}

\begin{figure*}
    \includegraphics[width=\linewidth]{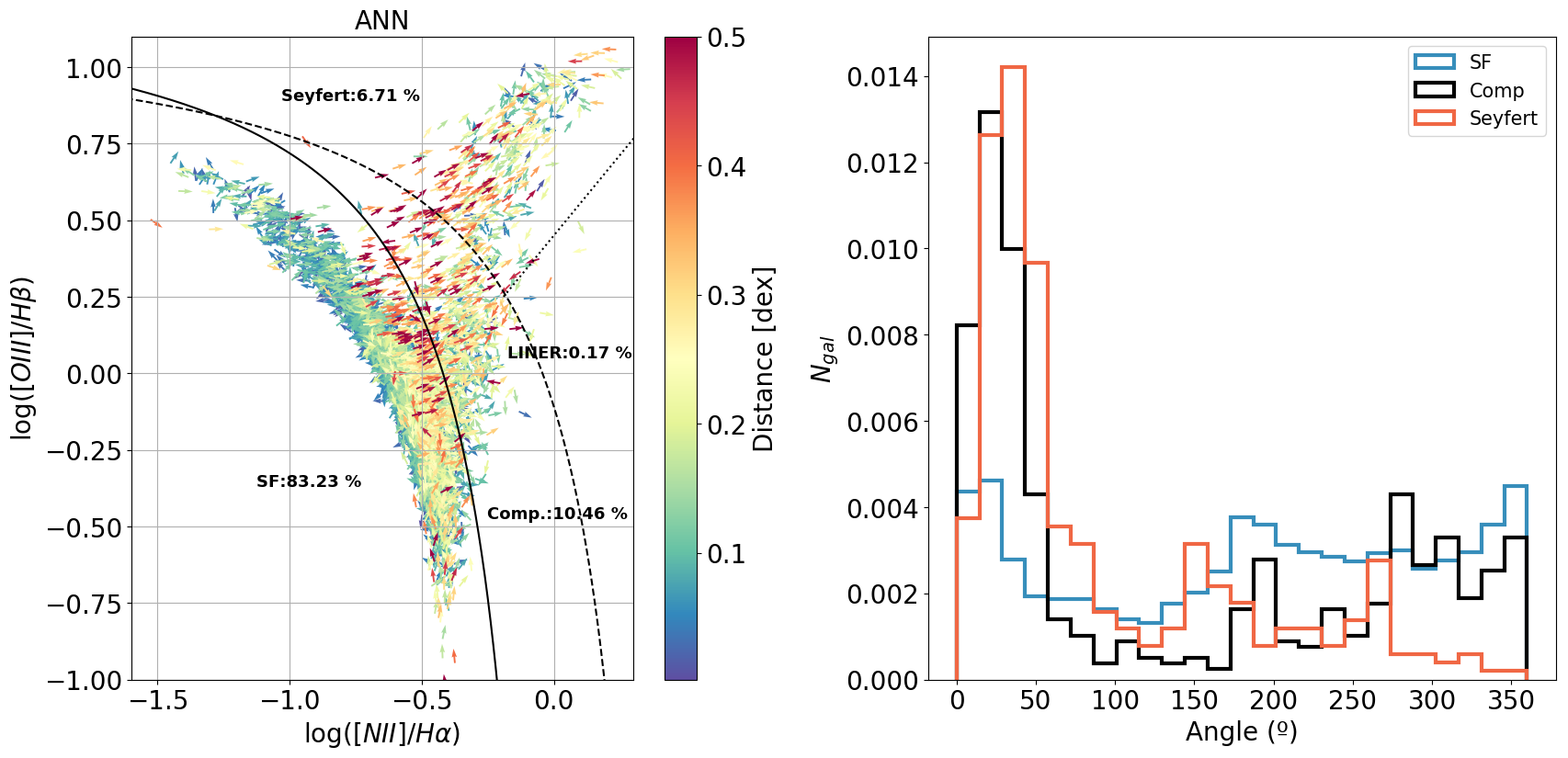}
 \caption{\tiny{ BPT diagram obtained by the ANN$_R$ trained with the CALMa set. Arrows point in the direction towards the location where galaxies should be placed according to their position in the SDSS MPA-JHU DR8 catalog. The color represents the distance for each point between the two BPT diagrams. The solid (ka03), dashed (Ke01) and dotted lines (S07) define the regions for the four main ionization mechanisms of galaxies. The percentage for each group is shown in black. The histograms on the rights represent the angular distribution of the arrows for Star forming, Seyfert and composite galaxies. The angle is defined as a clockwise rotation towards the $x$ axis.}}
\label{fig:BPT_dirrec}
\end{figure*}

In Fig. \ref{fig:BPT} we compare the BPT diagram recovered by the ANN$_R$ (left plot) and the one obtained from the SDSS testing sample (right plot). Galaxies are color-coded with the density of points and are grouped into four classes by three dividing lines: star-forming, composite, Seyfert, and LINER. The solid curve is derived empirically using the SDSS galaxies \citep[][hereafter ka03]{2003MNRAS.346.1055K}. The dashed curve is determined by using both stellar population synthesis models and photoionization \citep[][hereafter Ke01]{2001ApJ...556..121K}. The dotted line is a empirical division between Seyfert and LINER found by \citep[][hereafter S07]{2007MNRAS.382.1415S}. The sequence of metal enrichment experienced by star-forming galaxies from high to low values of the \oiii/\Hb ratio is clearly visible and well reproduced in the diagram. We will refer to that as the SF-wing. However, the saturation of the \nii/\Ha  and \oiii/\Hb ratios produces the migration of galaxies from right to left and from top to bottom lowering the percentage of Seyferts (from $10.04 \%$ to $6.71 \%$), composite (from $15.4 \%$ to $10.46 \%$) and LINERS galaxies (from $1.7 \%$ to $0.17 \%$) and increasing the percentage of star-forming galaxies (from $74.29 \%$ to $83.23 \%$).
\\\\ Other way to look at this is Fig. \ref{fig:BPT_dirrec}. We show the direction towards the location where galaxies should be placed in the BPT according to SDSS MPA-JHU DR8 catalog. The vectors are color-coded with the distance of each galaxy between the two BPT diagrams and more distance ones are plotted last. On average, star-forming galaxies deviate 0.10 dex while Seyfert and composite galaxies do 0.12 dex. On the right panel of Fig. \ref{fig:BPT_dirrec}, we plot the angular distribution of star-forming, Seyfert and composite galaxies. The angle is defined as a clockwise rotation towards the $x$ axis. While star forming galaxies do not show any preferential direction, Seyfert and composite galaxies point with an average angle of $45^o$ in the diagram. The CALMa set is very good at predicting the SF-wing because the main ionization mechanism in most of the regions in CALIFA and MaNGA galaxies is dominated by star-formation process. However, galaxies with high \nii/\Ha ratio are more difficult to constrain. 

%

\subsection{Comparison between different ANN$_R$ training sets}\label{subsec:table}
As we pointed out in the Sect. \ref{trreg} we have trained the ANN$_R$ with two different training samples. In the Appendix \ref{appSDSS} we show the results obtained with the SDSS training set in the SDSS testing sample. A quick look at these plots (Appendix \ref{linesSDSS}, \ref{ratiosSDSS} and \ref{fig:BPTSDSS}) proves the importance of testing the model on data with a different observational setup and calibration. Considering the fact that the EWs are estimated from a pseudo-spectrum (J-spectrum) with a much lower resolving power, the performance of the SDSS training set in SDSS testing sample is outstanding. Nevertheless, it would not be realistic to deduce from that the actual capability of this method to predict in J-PAS data. Testing the CALMa training set with SDSS galaxies or vice versa gave us a better picture of the weakness and inaccuracies of the model. For instance, the predictions made by ANN$_R$ trained with SDSS set on the \nii/\Ha and \oiii/\Hb ratios of MaNGA and CALIFA spaxels tend to be overestimated. This is the opposite effect observed when the ANN$_R$ is trained with CALMa training set and tested on SDSS galaxies. The performance on the validation samples, that is, the data that belongs to the same survey, is generally better. \\\\ For the sake of illustrating the performance of both training sample (SDSS test and CALMa set) in each one of the testing sets (CALIFA, MaNGA and SDSS) we create a comparison table (Table \ref{tablebias}). As it can observe, there will always be a line that is better recovered in one particular simulation, for example \Ha in CALMa vs SDSS, but the overall performance of the ANN$_R$ is generally more accurate with data from the same survey. 
\begin{table*}[htb]
 \tiny
 \centering
\begin{tabular}{|l|l|l|l|l|l|l|l|l|}
\hline
Training vs Test &  \Ha $ (\%)$  & \Hb $ (\%)$  & \oiii $ (\%)$ & \nii $ (\%)$ & \nii/\Ha [dex] & \oiii/\Hb [dex]  & $ \ion{O}{3}\ion{N}{2}$ [dex] \\ \hline
SDSS vs SDSS     &  $-0.4 \pm 8.3$& $-2.3 \pm 12.3$  & $1.8 \pm 15.9$ & $2.6 \pm 16.3$  &   $0.019 \pm 0.089$ & $0.026 \pm 0.080$ & $0.012 \pm 0.119$ \\ \hline
SDSS vs CALIFA     &  $-6.3 \pm 10.7$& $-12.5 \pm 13.5$  & $-5.3 \pm 21.1$ & $-2.3 \pm 21.4$  &  $0.018 \pm 0.122$ & $0.04 \pm 0.102$ & $0.023\pm 0.159$ \\ \hline
SDSS vs MaNGA    &  $-2.4 \pm 11.1$& $-8.1 \pm 13.7$  & $-3.5 \pm 19.9$ & $9.8\pm 22.1$  &   $0.06 \pm 0.105$ & $0.033 \pm 0.096$ & $-0.031 \pm 0.148$ \\ \hline
CALMa vs CALIFA     &  $-4.4 \pm 8.1$& $-4.9 \pm 12.2$  & $1.5 \pm 19.2$ & $-3.8\pm 15.3$  &  $0.003\pm 0.088$ & $0.035 \pm 0.089$ & $0.037 \pm 0.131$ \\ \hline
CALMa vs MaNGA    &  $-2.3 \pm 8.6$& $-1.7 \pm 12.2$  & $0.4 \pm 17.4$ & $8.4\pm 18.2$  &   $0.051\pm 0.083$ & $0.019 \pm 0.077$ & $-0.03 \pm 0.125$ \\ \hline
CALMa vs SDSS     &  $0.2 \pm 8.7$& $5.4 \pm 14.3$  & $4.8 \pm 16.4$ & $-6.4\pm 15.9$  &  $-0.028\pm 0.102$ & $-0.004 \pm 0.091$ & $0.038 \pm 0.112$ \\ \hline
\end{tabular}
\caption{\tiny{Relative difference between the EWs (in percentage) and ratios (in dex) predicted by ANN$_R$ and the values provided by the testing samples. The comparison is made between the training sample proposed in this paper and SDSS, CALIFA and MaNGA testing sample.}}
\label{tablebias}
\end{table*}

 \subsection{The 5max method in practice}\label{subsec:test5max}
\begin{figure}[htb]
 \centering
 \includegraphics[width=\hsize]{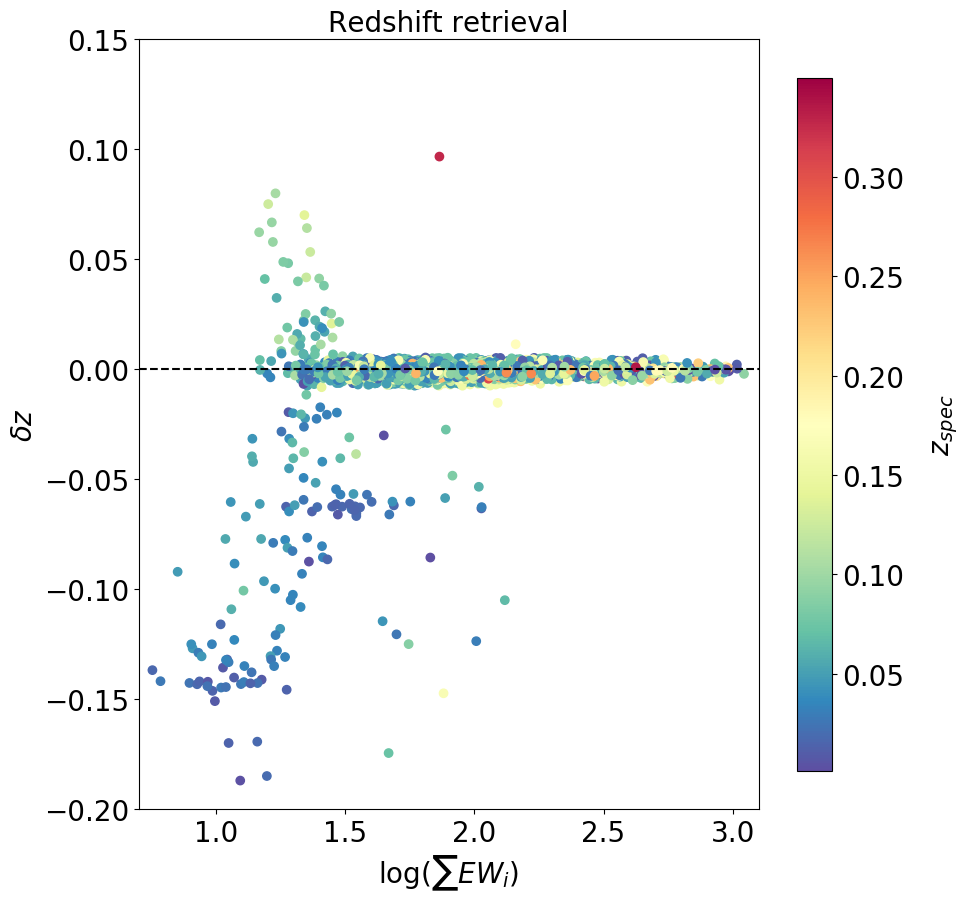}
  \caption{\tiny{$\delta z$ obtained from the difference between the spectroscopic redshift and the median redshift in the \textit{5max} setting in function of the sum of the EWs provided in the SDSS catalog for a total of 10000 galaxies. Points are color-coded with the spectroscopic redshift.}}
 \label{figzre}
 \end{figure}

A simple test to confirm the capability of the \textit{5max} method to retrieve the redshift of the object is to verify whether the average redshift over the five configuration is far from the true redshift. Normally, we would compute the EWs only in the redshift within the PDF of photo-zs before applying the \textit{5max}, but let us assume we do not have any information regarding the redshift of the object. Then, we have to calculate the EWs in all the redshift from $0$ to $0.35$ inside the grid and pick only the five redshifts that maximize their sum. Fig. \ref{figzre} shows this scenario where points are color-coded with the spectroscopic redshift. For emission line galaxies ($\sum EW_i > 20 $ \AA), this method is able to obtain the redshift of the object with high precision; what is more, the redshift is not needed as an input. Nevertheless, the \textit{5max} is not able to retrieve the redshift of the object when galaxies have low emission. The set of redshifts that maximizes the sum of the EWs is largely uncertain and consequently we do need the PDFs to constrain the redshift value. 
 \subsection{Dependency on the equivalent width and redshift uncertainty}\label{subsec:EWdependence}
\begin{figure}[htb]
 \centering
 \includegraphics[width=\hsize]{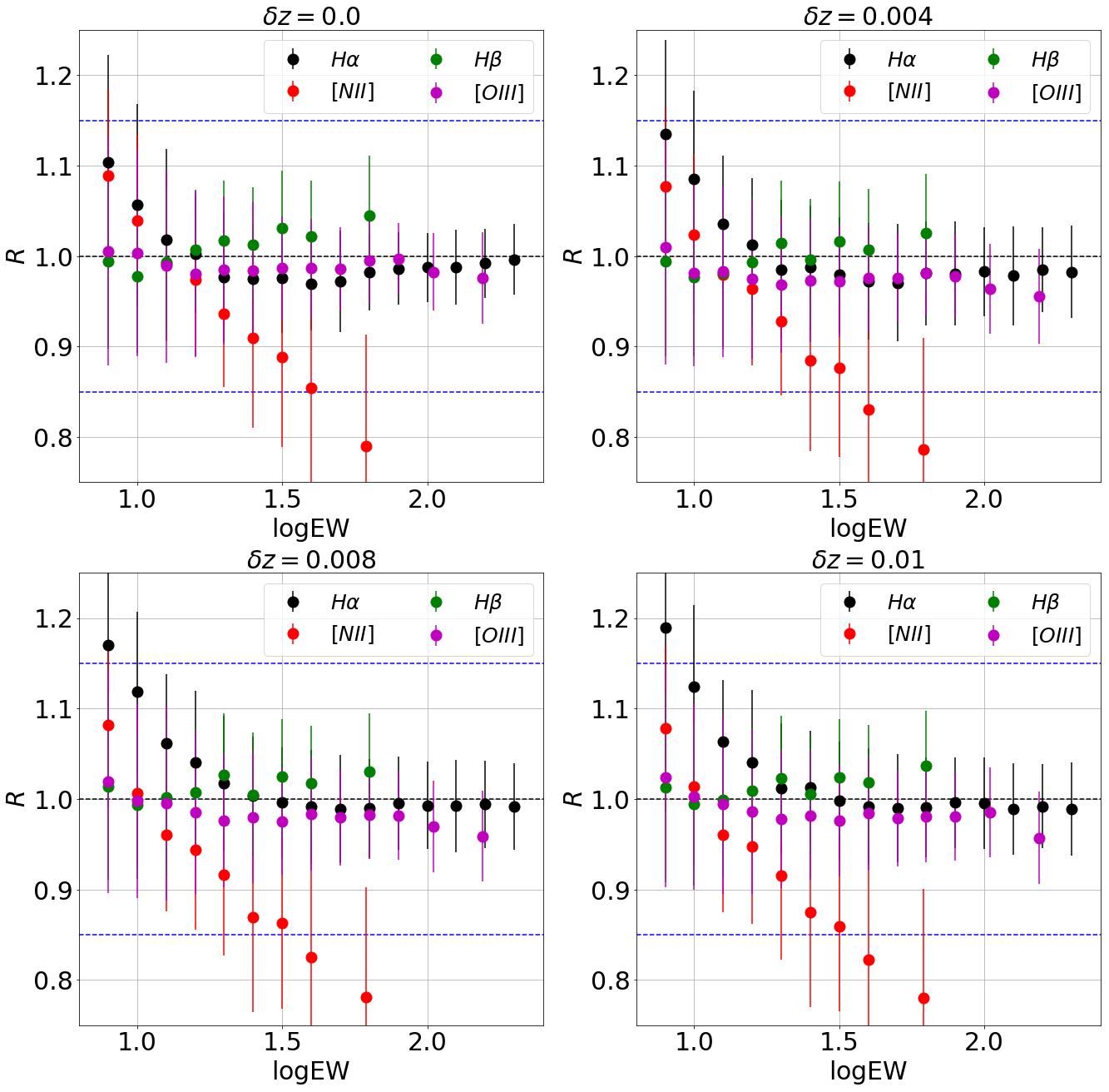}
  \caption{\tiny{Each point represents the median ratio between the predicted and the observed SDSS EWs and bars indicate the mean absolute deviation. Each bin contains 500 galaxies  in the interval $10^\gamma < EW_{SDSS} < 10^{\gamma+0.1}$ with $\gamma$ ranging from 0.8 to 2.5 for \Ha,  from 0.8 to $2.2$ for \oiii,  from 0.8 to $1.8$ for \Hb and  from 0.8 to $1.8$ for \nii. From left to right and top to bottom we increase the uncertainty in the redshift. Dashed blue lines point to a ratio of 1.15 and 0.85 respectively. Dash black line represent zero bias between the predicted and observed EWs.}}
 \label{fig:EWdependecy}
 \end{figure}

In order to explore the limitation of the model as a function of the redshift uncertainty and the EW of each one of the emission lines, we assemble galaxies in bins by the EW provided in the SDSS catalog and compute the ratio $(R)$ between the predicted and observed EW. Each bin contains 500 galaxies in the interval $10^\gamma < EW_{SDSS} < 10^{\gamma+0.1}$ with $\gamma$ ranging from 0.8 to 2.5 for \Ha,  from 0.8 to $2.2$ for \oiii,  from 0.8 to $1.8$ for \Hb and  from 0.8 to $1.8$ for \nii. As we observe in Fig. \ref{fig:EWdependecy}, \Ha is clearly more affected by the \textit{5max} strategy when $EW(\Ha) \le 10^{1.2}$ \AA. Independently of the redshift uncertainty, the ANN$_R$ trained with the CALMa set has more difficulties to constrain the \nii line underestimating its value as the EW increases. It also presents more dispersion, which is an indication of the different galaxy population found at such EW bins. This is, the percentage of galaxies hosting an AGN is higher. 
Nonetheless, we are able to constrain the EW of galaxies with a bias less than $10 \%$ for most of the lines even with high uncertainty in the redshift. 

 \subsection{EW limit}\label{subsec:EWlimit}
\begin{figure}[htb]
 \centering
 \includegraphics[width=\hsize]{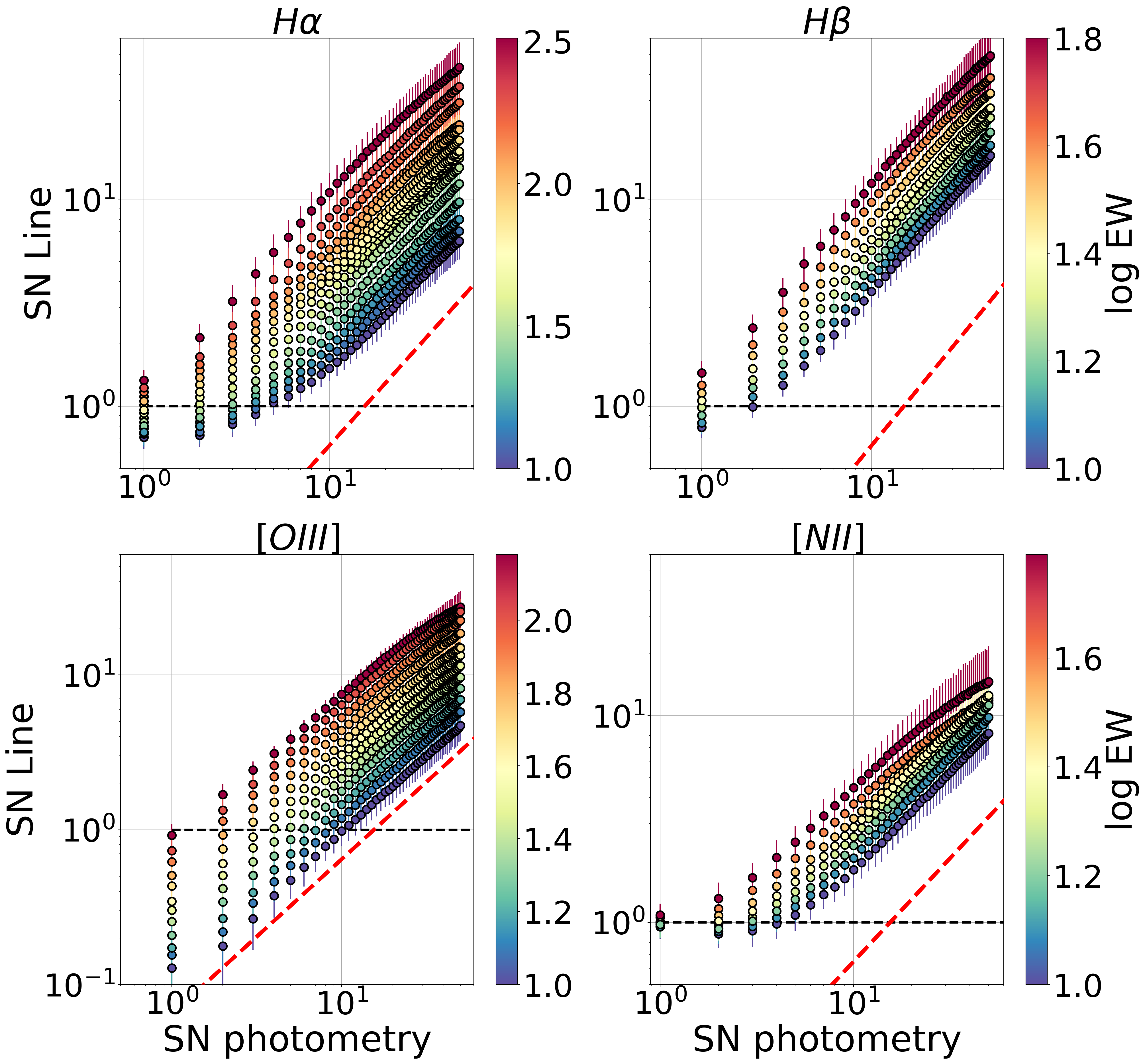}
  \caption{\tiny{Predicted SN of \Ha, \Hb, \oiii and \nii lines in function of the SN in the photometry. For a given SN in the photometry, each point represent the mean SN obtained in the line for 500 SDSS galaxies in the interval (color-coded) $\gamma < \log EW_{SDSS} < \gamma+0.1$ with $\gamma$ ranging from 0.8 to 2.5 for \Ha,  from 0.8 to $2.2$ for \oiii,  from 0.8 to $1.8$ for \Hb and  from 0.8 to $1.8$ for \nii. Errors bars indicate the mean absolute deviation. Dashed red line represents Eq. \ref{Eq:SN} for $EW = 10$ \AA.}}
 \label{Fig:SN_analysis}
 \end{figure}

The minimum EW measurable in a photometry system using a traditional method depends only on the SN of the photometry and the effective width of filters in the system. Let us assume, that an emission line falls within one filter ($f_{i}$) and we know with high precision the redshift of the object. The EW of an emission line can be computed assuming the line is infinitely thin as:
 \begin{equation}
EW = \Delta' (\lambda_z)(Q - 1)
\end{equation}
 where $\Delta' $ is the effective width of filter $f_{i}$ and $Q$ is the ratio between the flux with and without emission line see \citep[see][for details]{2007PASP..119...30P} or simply:
\begin{equation}
Q = 10^{-(m^{obs}_{AB} - m^{cont}_{AB})/2.5}
\end{equation}
 in AB magnitudes. Then, if we are able to estimate the flux of the stellar continuum in the filter tracing the emission line, obtaining the EW is straightforward. The SN of such line can be expressed in terms of Q and the SN of the photometry in the filter $f_{i}$ through the following equation:
\begin{equation}
SN_{EW} = \frac{Q-1}{Q} SN_{phot}
\label{Eq:SN}
\end{equation}
 The minimum EW measurable can be written as:
\begin{equation}
EW_{min} = \frac{\Delta '}{SN_{phot} - 1}
\label{Eq:EWmin}
\end{equation}
For SN$_{phot} = 10$ only lines with EW greater that $16.1$ \AA $ $ can be measured in a filter width of $145$ \AA $ $. 
\\\\In Fig. \ref{Fig:SN_analysis} we determine the relation between the SN of each line obtained with the ANN in function of the SN of the photometry. As before, we assume no errors in the redshift of the objects. We analyze here the same galaxies used in the previous section in order to study the dependence with the EW. Each color represents the average SN obtained in the line for 500 SDSS galaxies with the same EW. The red dashed line follows Eq. \ref{Eq:SN} for $EW = 10$ \AA, $ $ which is the lowest EW bin considered in the simulations. All the lines estimated with the ANN can be measured with a precision higher than a method based on the contrast between the emission line flux and the stellar continuum. 
\\\\ \Hb is the line that can be obtained with the highest SN for the same EW, even with better precision than \Ha. This is not surprising since the algorithm has learnt the implicit relation between \Ha and \Hb constrained by the Balmer series and the amount of interstellar dust. Therefore, an EW in \Hb of 10 \AA, that corresponds in average to an EW in \Ha of about 30 \AA,  are measured with the same SN. More complex relations such as the one between \Ha and \nii has also been learnt, but we observe a flattening of the SN of the \nii line for the highest EW with an increase in the scatter. This regime is populated with more AGN-like galaxies and consequently it is more difficult to constrain with the CALMa set. This finding agrees with the behaviour observed in Fig. \ref{fig:EWdependecy}, where higher values of \nii are systematically underestimated. Finally, the \oiii line is generally more difficult to constrain as we obtain lower SN ratio. Nevertheless, it can be recovered with better precision than a method based only on the photometry contrast.
\\\\ To sum up, with an ANN one can measure a EW of 10 \AA $ $ in \Ha, \Hb, \nii and \oiii lines with a SN in the photometry of  5, 1.5, 3.5 and, 10 respectively. However, methods based on the photometry contrast need for the same EW a SN in the photometry of at least 15.5. These facts illustrate once again the capability of machine learning algorithms to go beyond in precision and accuracy respect to traditional methods when large amount of data sets are available.

\section{Comparison between miniJPAS and SDSS}\label{aegisSDSS}
In this section we analyze and compare the data from the SDSS survey that has also been observed with miniJPAS in the AEGIS field. Firstly, we describe the miniJPAS survey in Sect. \ref{subsect:miniJPAS}. Then, we analyze and compare the properties of galaxies in terms of their emission lines in Sect. \ref{subsect:miniJPASvsSDSS}. 
\subsection{miniJPAS survey}\label{subsect:miniJPAS}
The miniJPAS survey \citep{2020arXiv200701910B} is the result of the J-PAS-Pathfinder observation phase carried out with the $2.55$ m telescope (T250) at the Observatorio Astrofísico de Javalambre in Teruel (Spain). miniJPAS was observed with the Pathfinder camera, the first instrument installed in the T250 before the arrival of the Javalambre Panoramic Camera \citep[JPCam, ][]{2019A&A...622A.176C,2014JAI.....350010T,2015IAUGA..2257381M}. JPAS-\textit{Pathfinder} instrument is a single CCD direct imager ($9.2k \times 9.2k$, $10\mu m$ pixel) located at the center of the T250 FoV with a pixel scale of 0.23 arcsec pix$
^{-1}$, that is vignetted on its periphery, providing an effective FoV of 0.27 deg$^2$. The miniJPAS data includes four pointings of 1 deg$^2$ in total along the Extended Groth Strip (called the AEGIS field). We use the same photometric system of J-PAS. Thus, AEGIS was observed with 56 narrow band filters covering from $\sim 3400$ to $\sim 9400$ \AA $ $. Observations in the four broad bands ($u_{JPAS}$, and SDSS g, r, and i) were also taken. More than 60000 objects were detected in the r band, allowing to build a complete sample of extended sources up to $r \le 22.7$ (AB). A detailed description of the survey is in \cite{2020arXiv200701910B}. Data is accessible and open to the community through the web page of the survey\footnote{http://www.j-pas.org/}.
\subsection{miniJPAS vs SDSS}\label{subsect:miniJPASvsSDSS}
\begin{figure}
\includegraphics[width=0.24\textwidth]{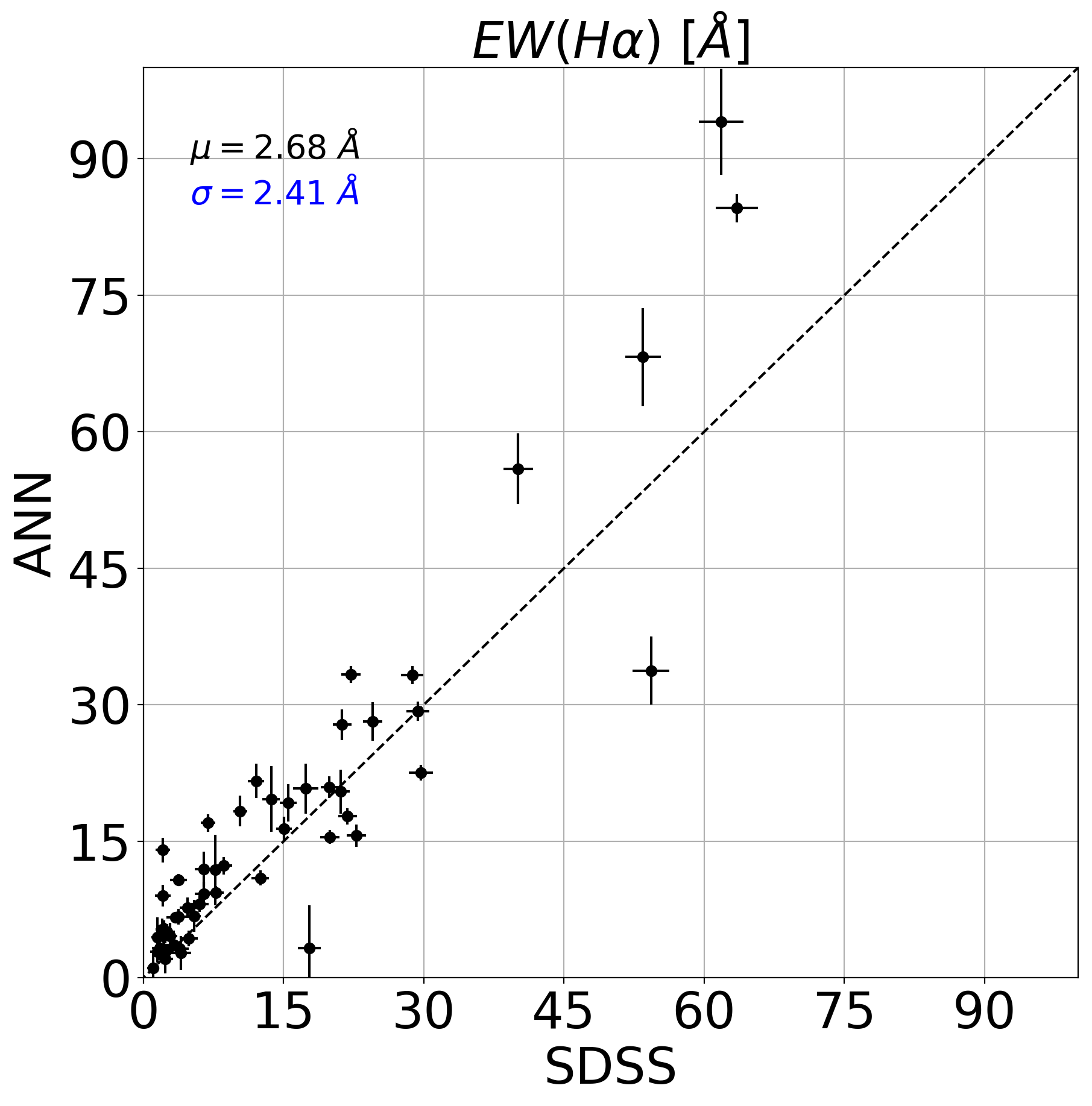}
\includegraphics[width=0.24\textwidth]{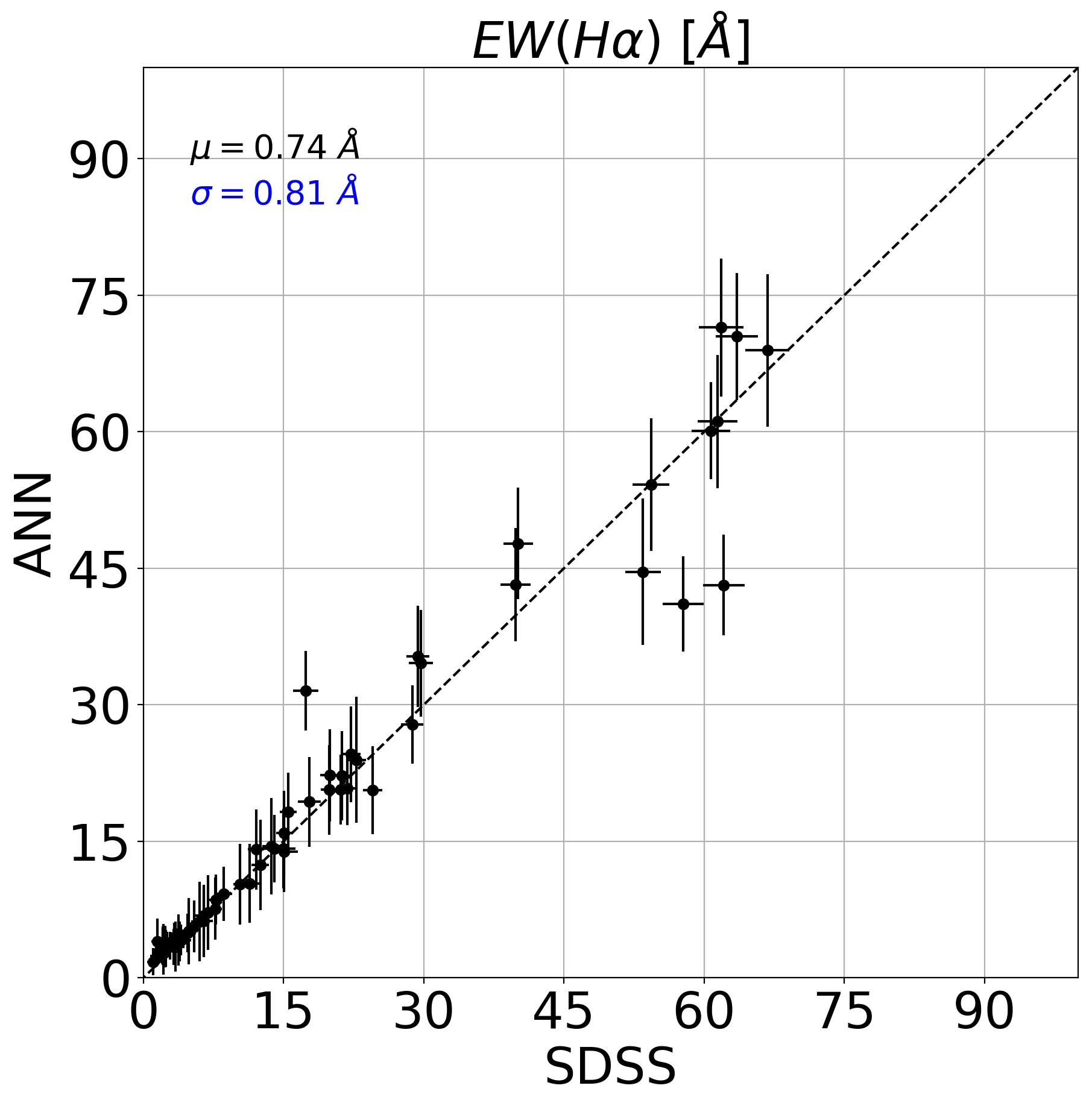}

\includegraphics[width=0.24\textwidth]{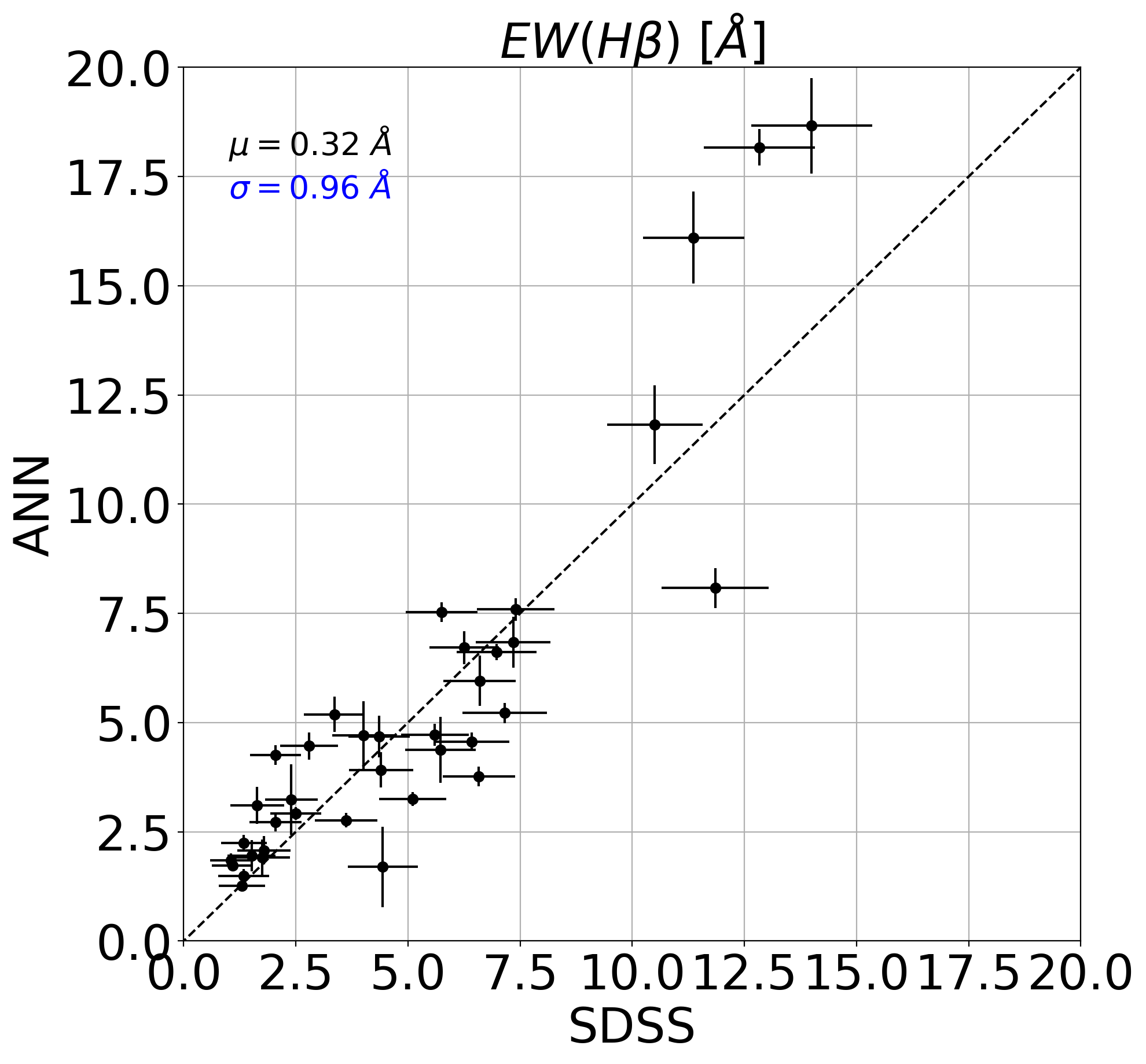}
\includegraphics[width=0.24\textwidth]{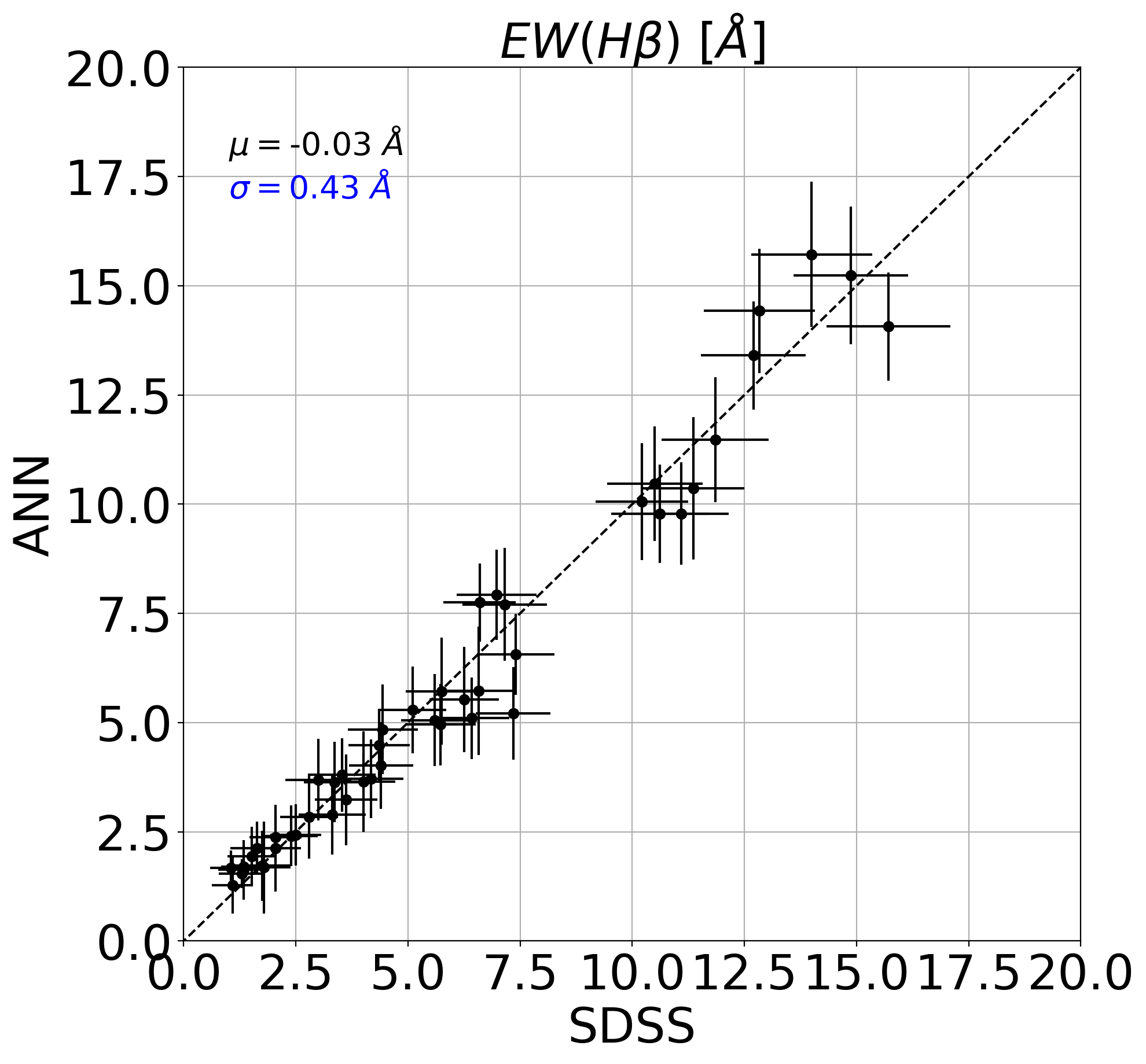}

\includegraphics[width=0.24\textwidth]{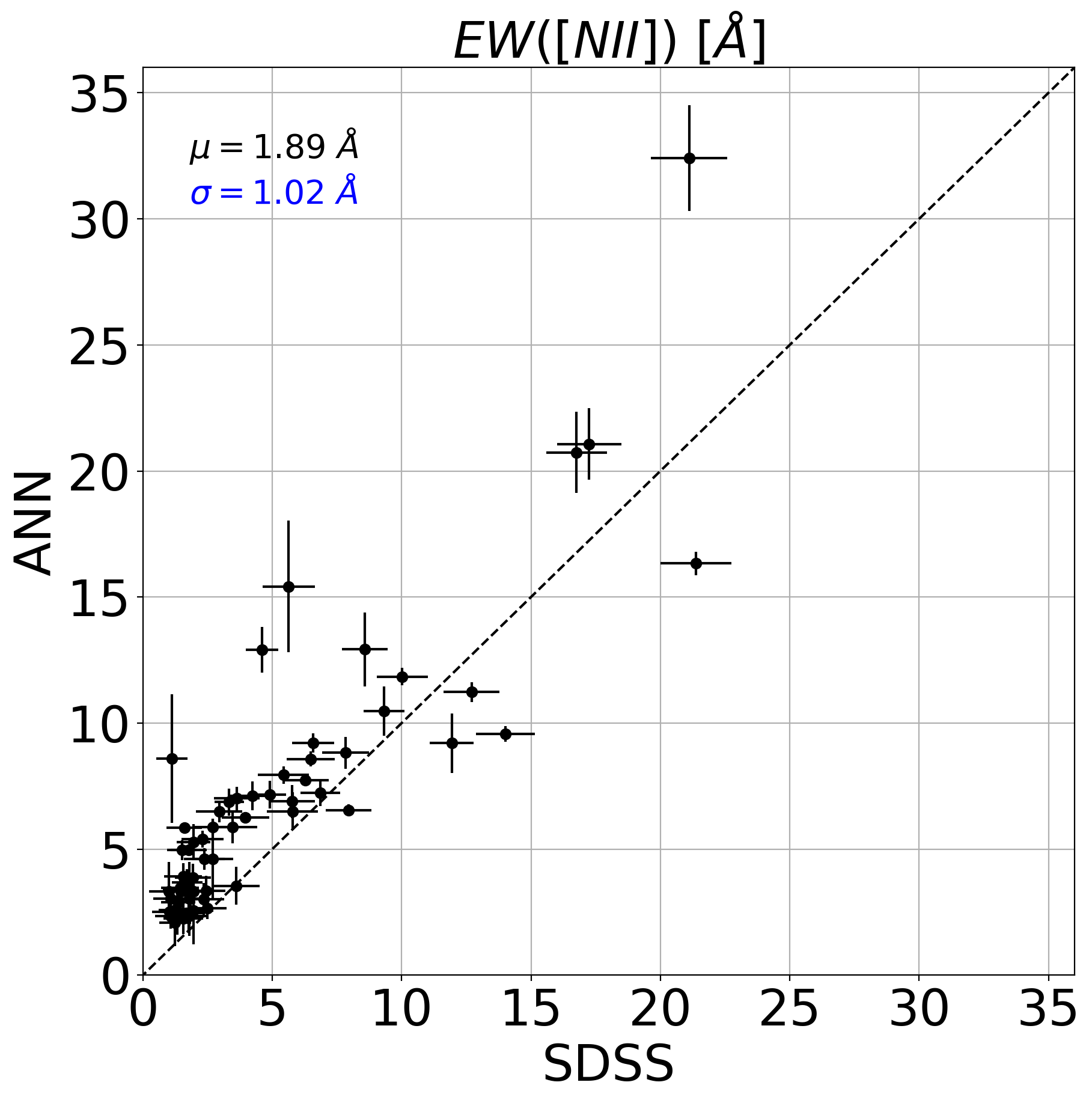}
\includegraphics[width=0.24\textwidth]{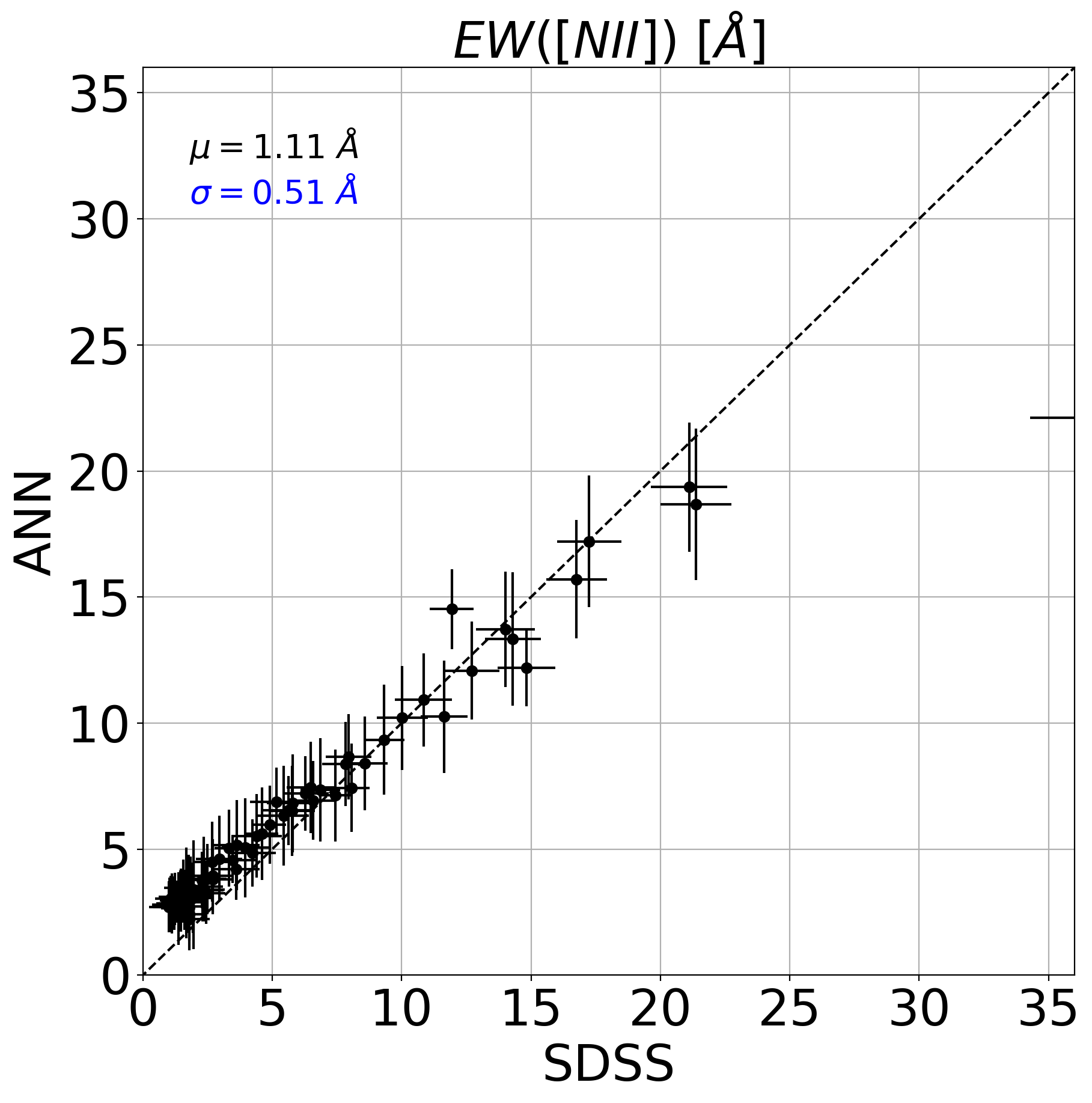}

\includegraphics[width=0.24\textwidth]{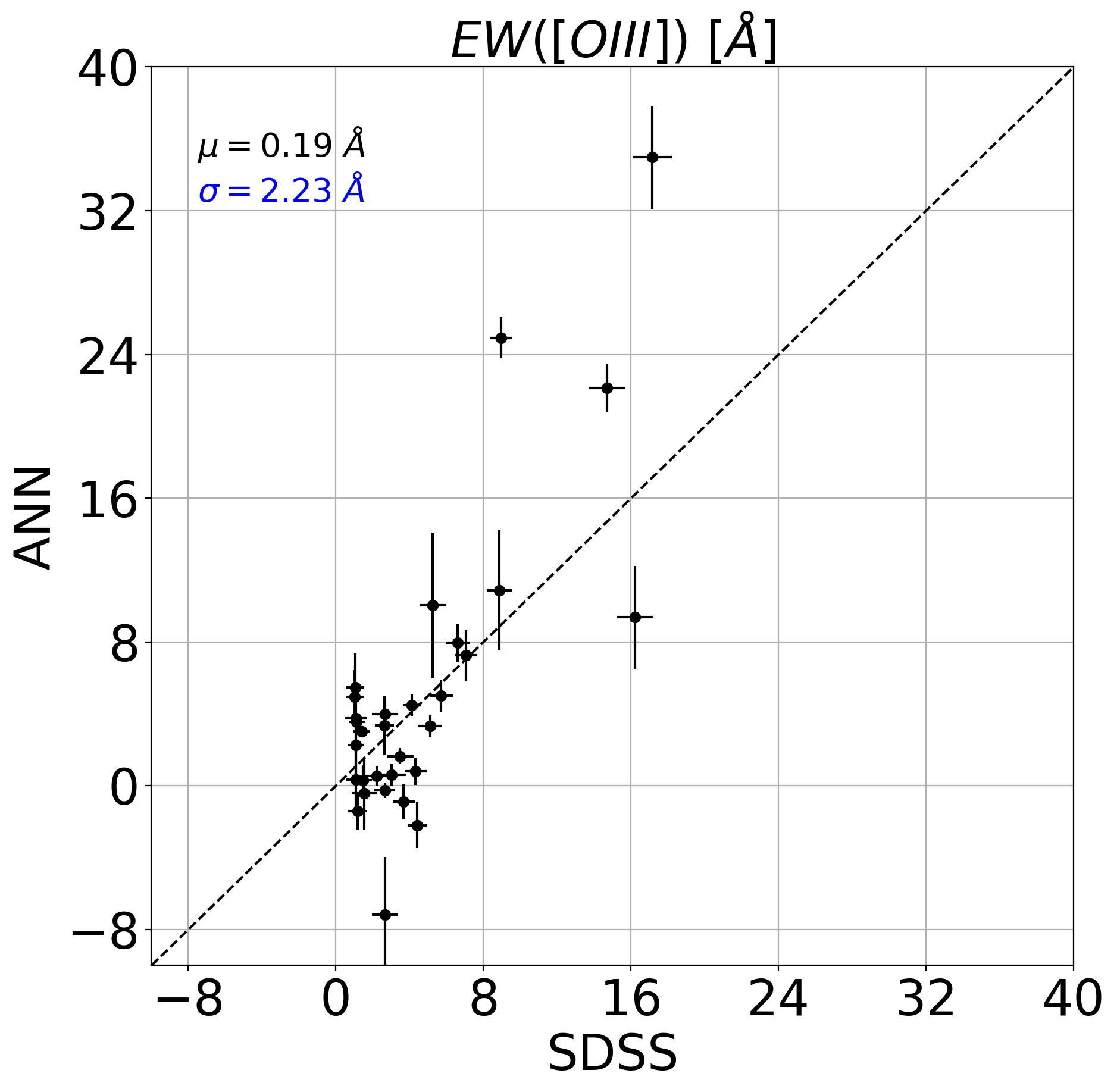}
\includegraphics[width=0.24\textwidth]{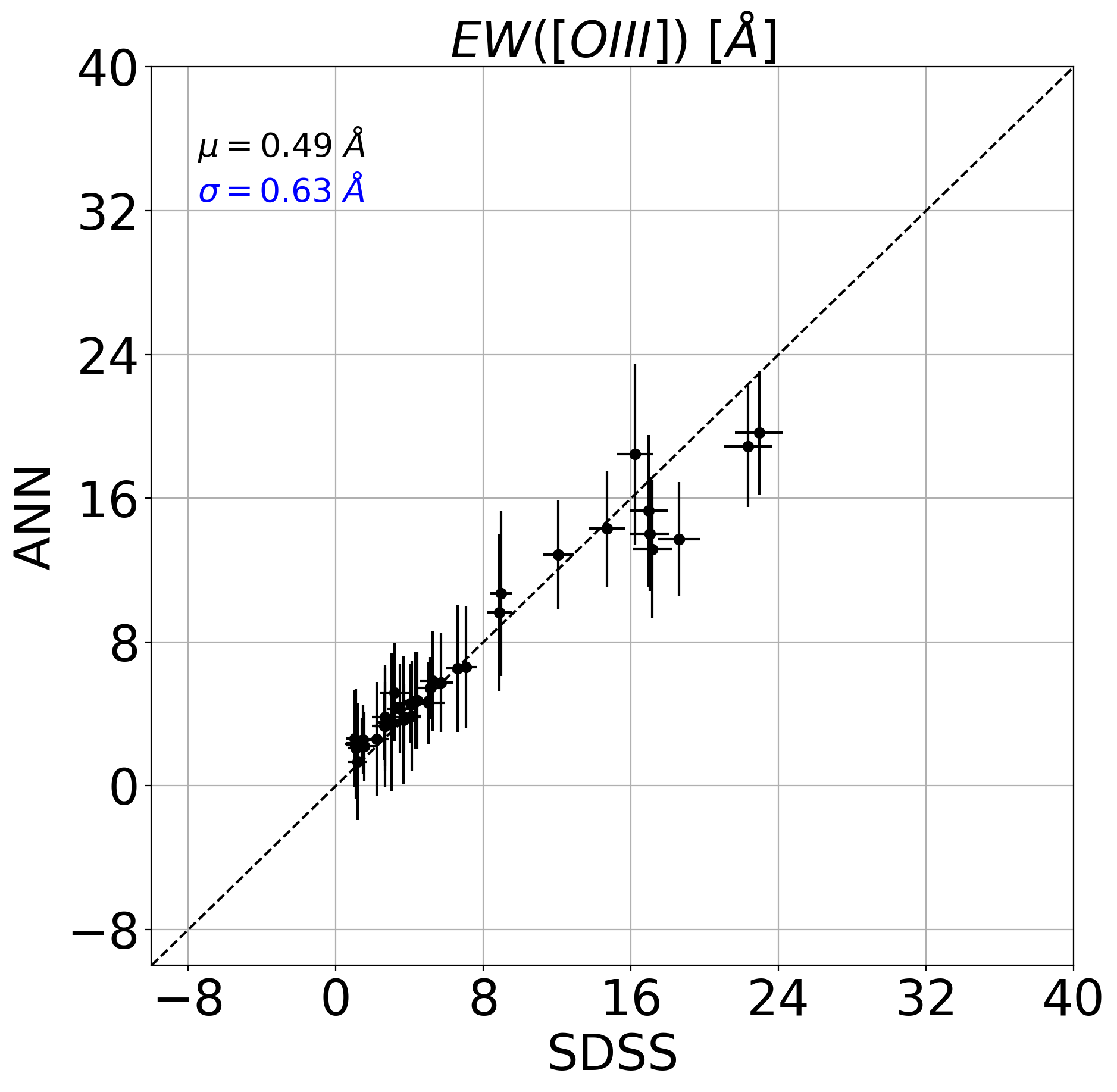}
\caption{\tiny{Comparison between the EWs of \Ha, \nii, \Hb  and \oiii  measured  in  the  SDSS  spectra  and  the  predictions made by the ANN on miniJPAS data using the \texttt{MAG} \texttt{PSFCOR} (left panel) and synthetic J-PAS magnitudes obtained from the SDSS spectra (right panel). Black and blue numbers are the median and the median absolute deviation of the difference. Dashed black line is line with slope one.}}
\label{figAEGISSDSS}
\end{figure}

\begin{figure*}
	\includegraphics[width=6.4cm,height=6cm]{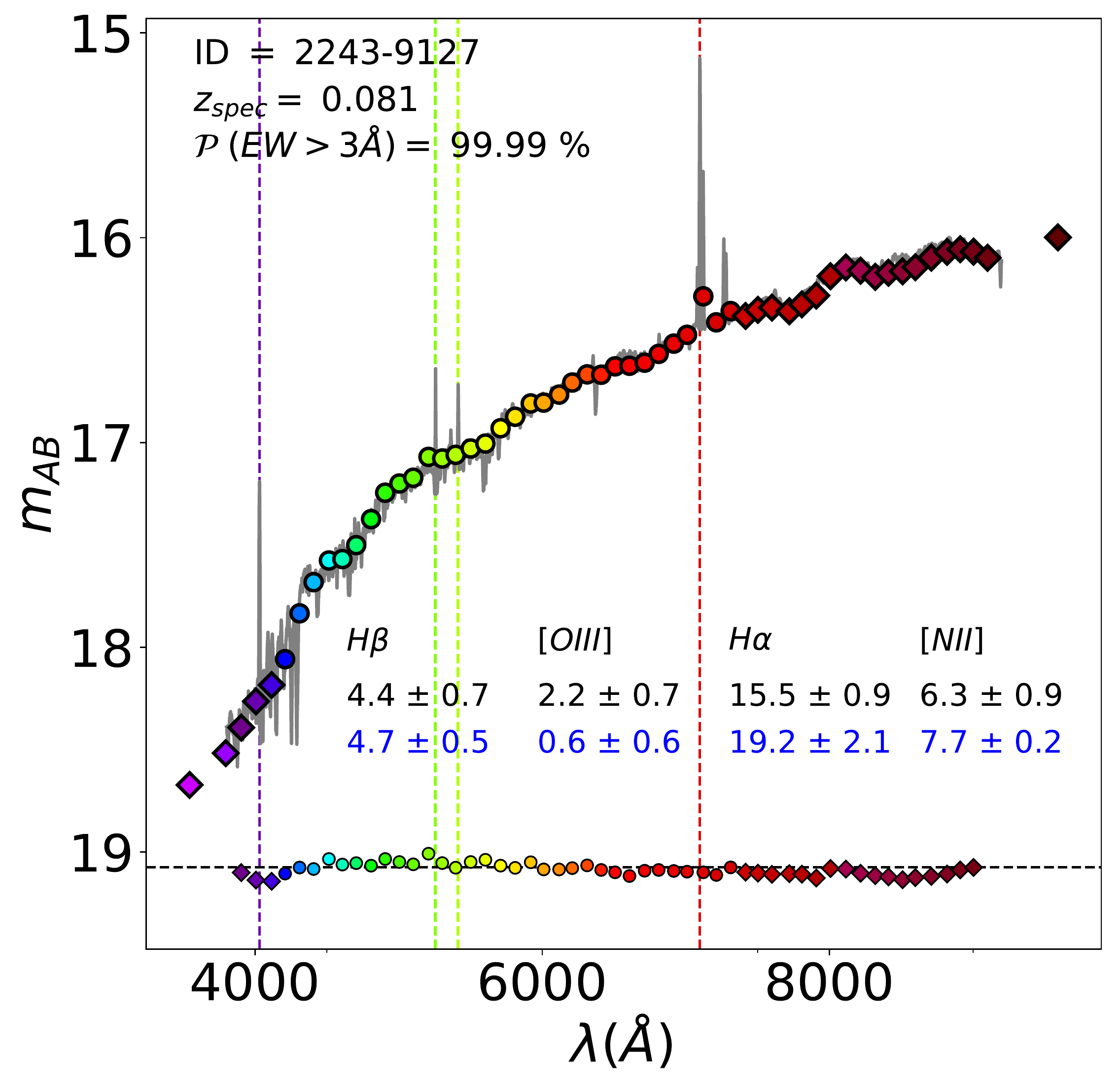}
	\includegraphics[width=6.4cm,height=6cm]{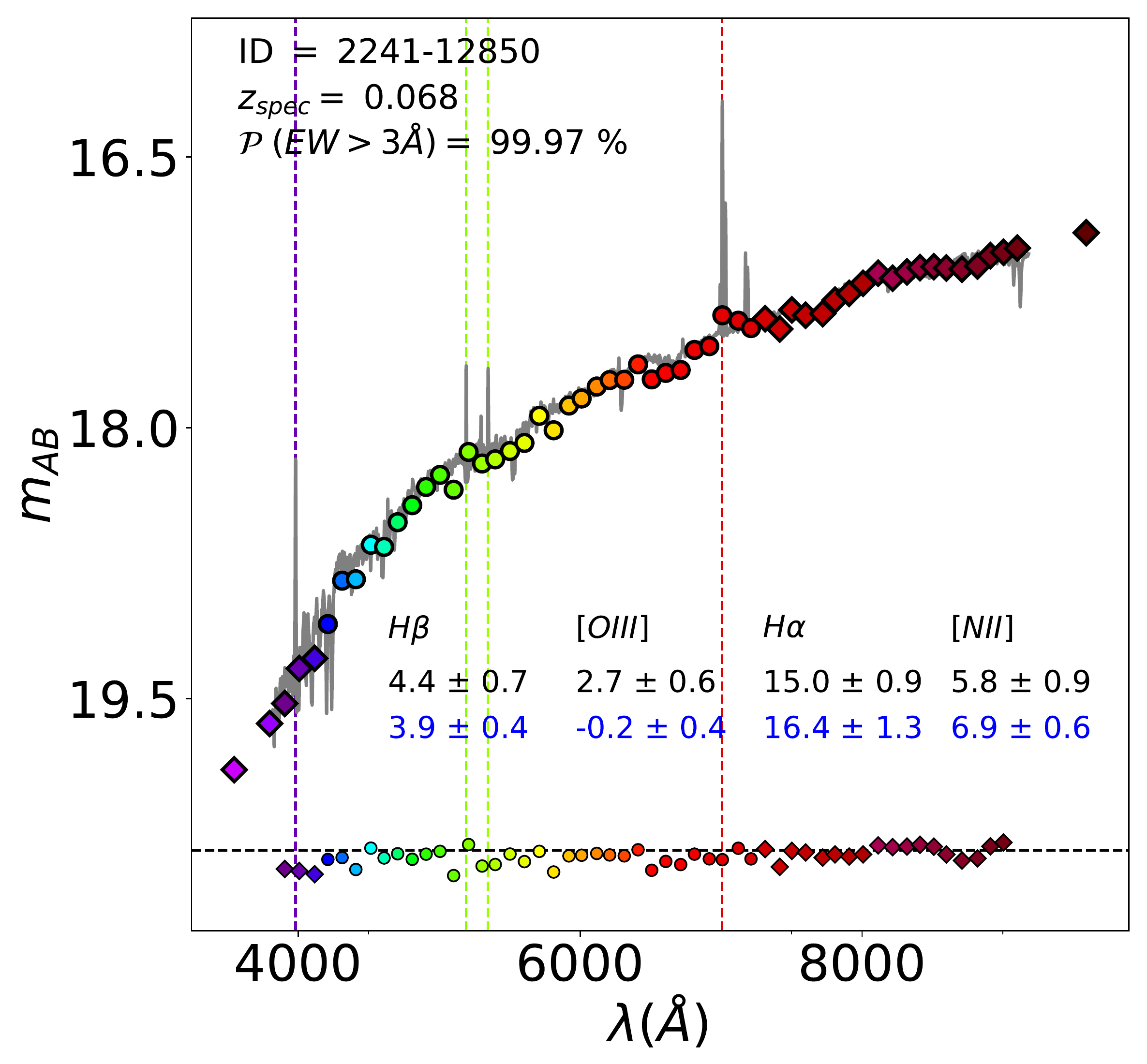}
	\includegraphics[width=6.4cm,height=6cm]{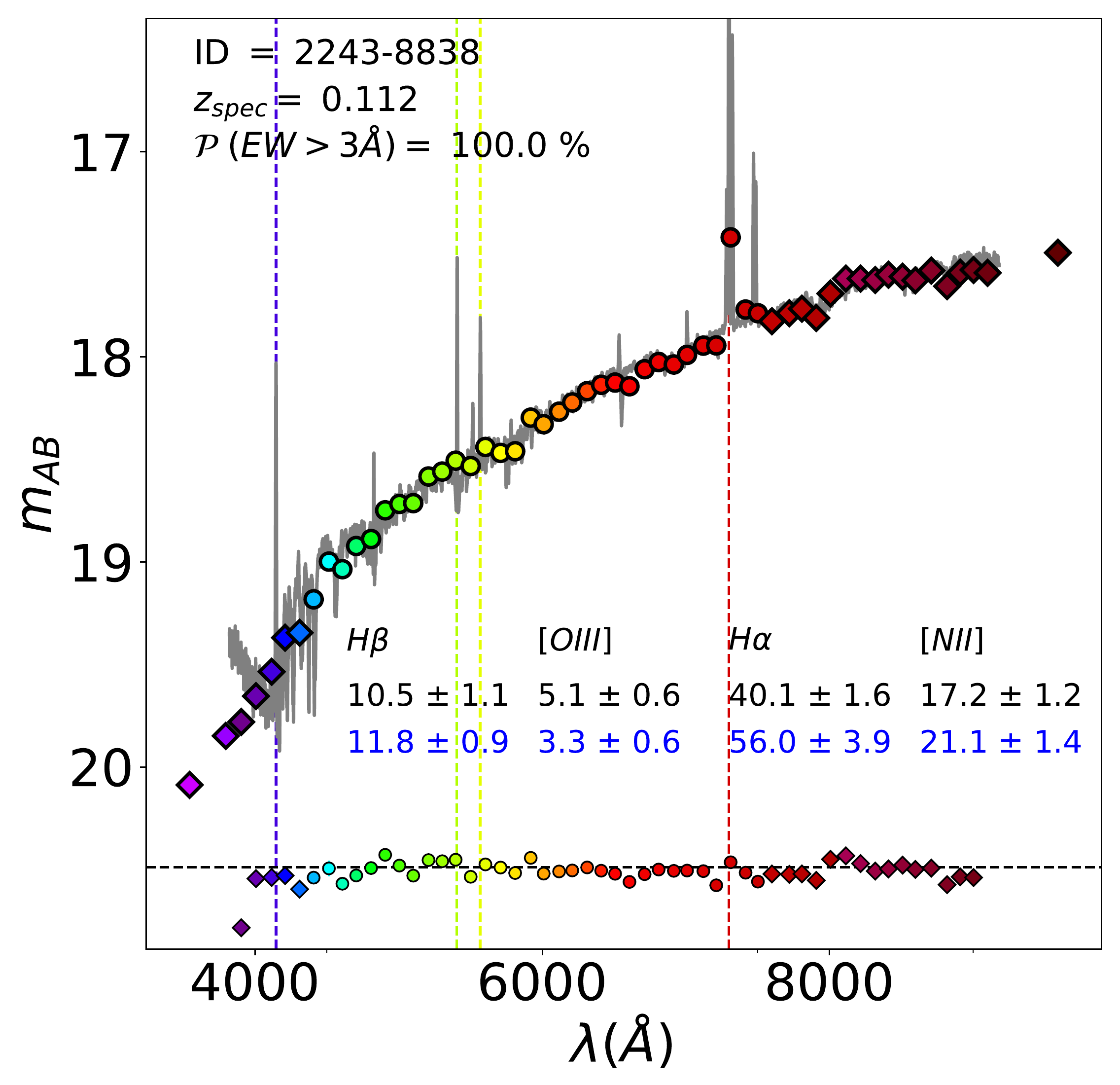}

	\includegraphics[width=6.4cm,height=6cm]{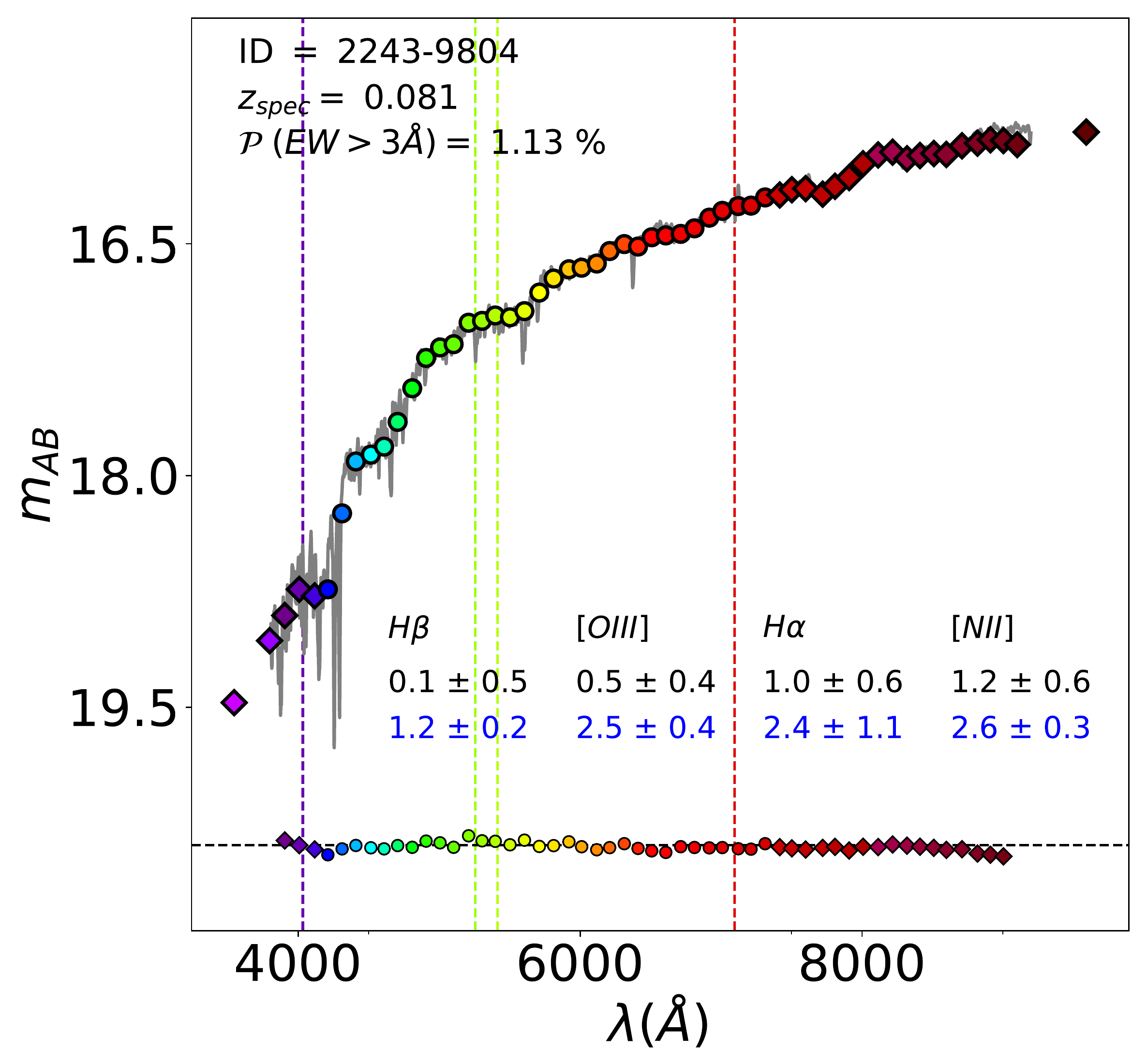}
	\includegraphics[width=6.4cm,height=6cm]{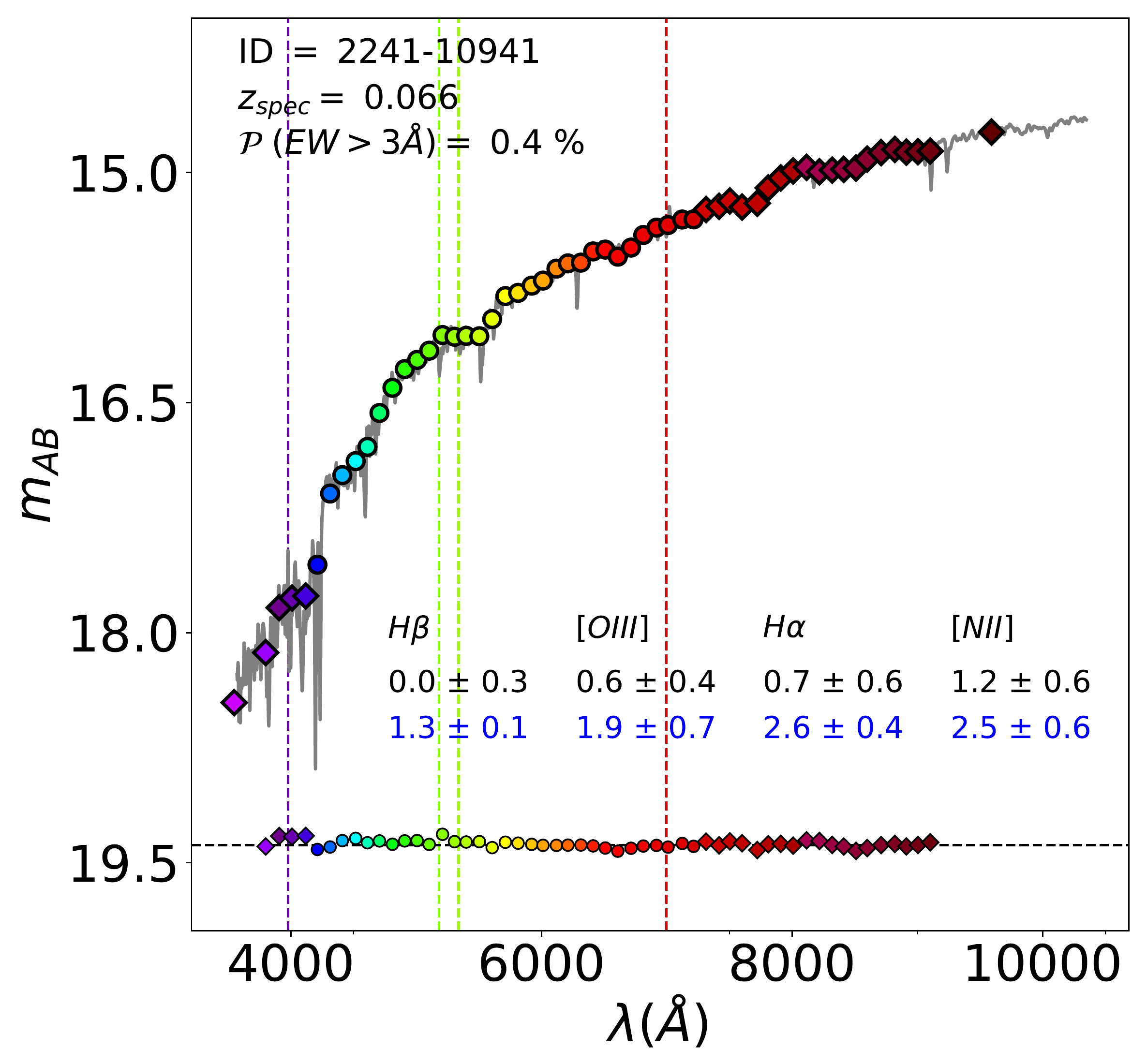}
	\includegraphics[width=6.4cm,height=6cm]{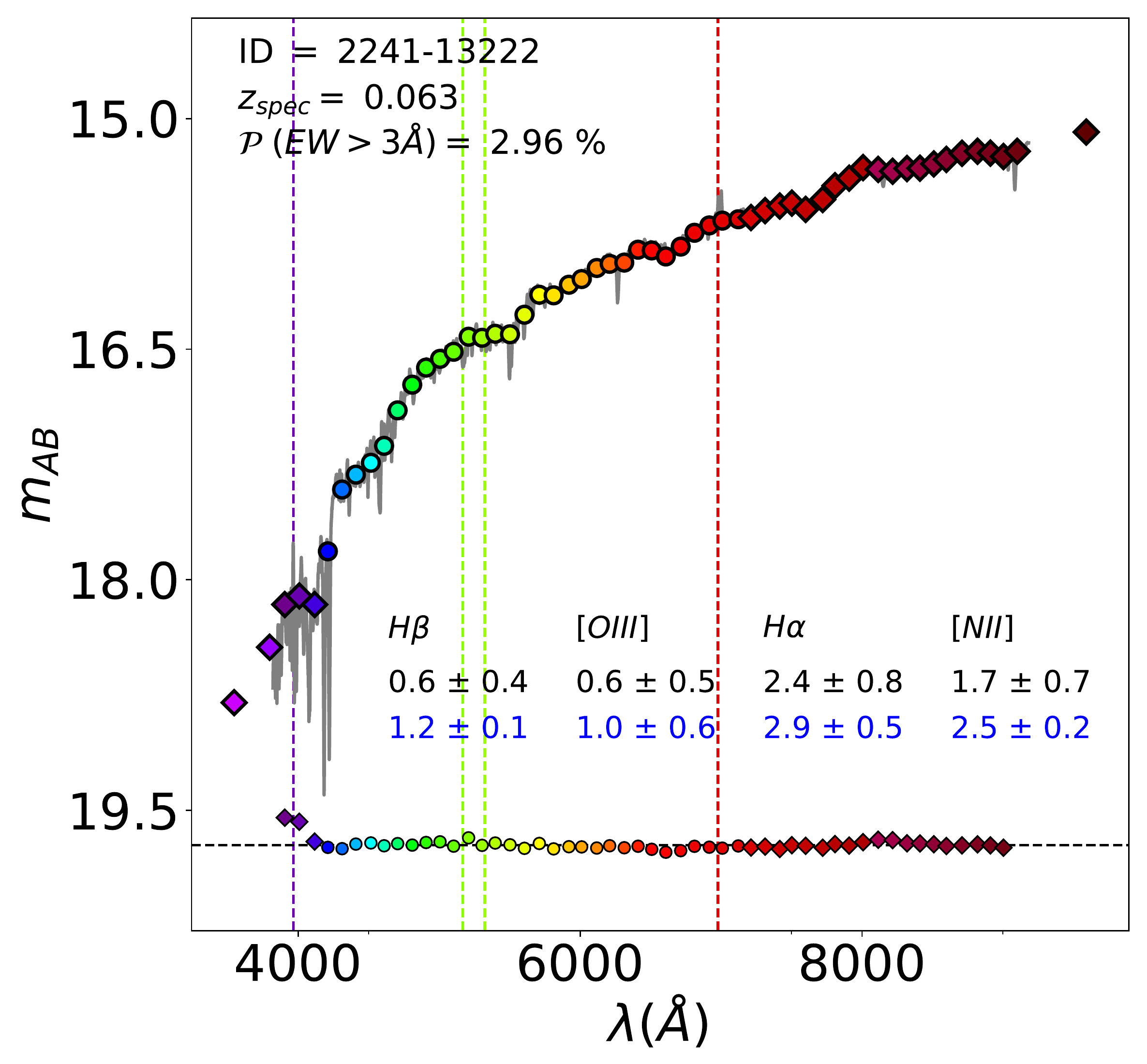}

	\includegraphics[width=6.4cm,height=6cm]{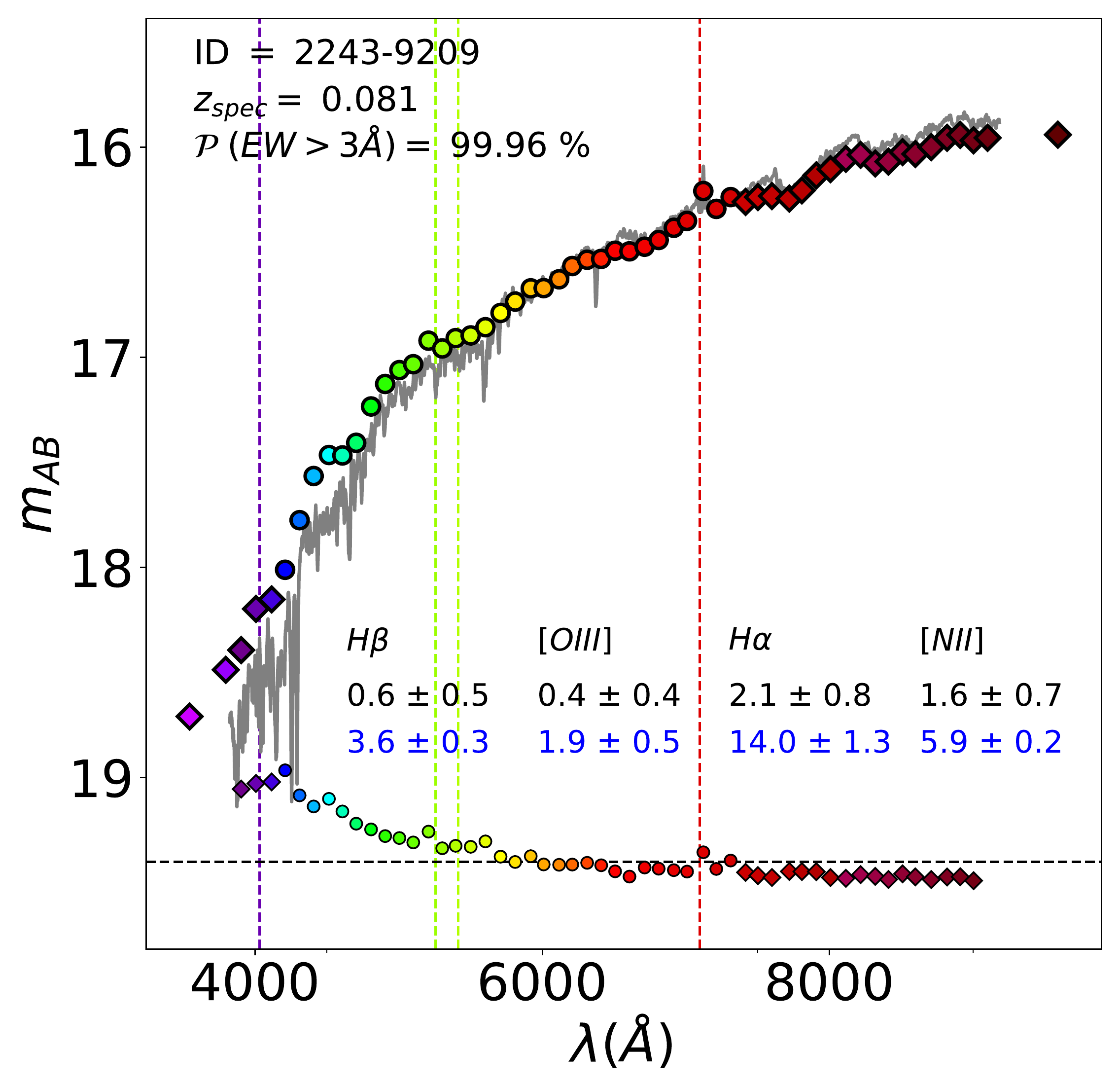}
	\includegraphics[width=6.4cm,height=6cm]{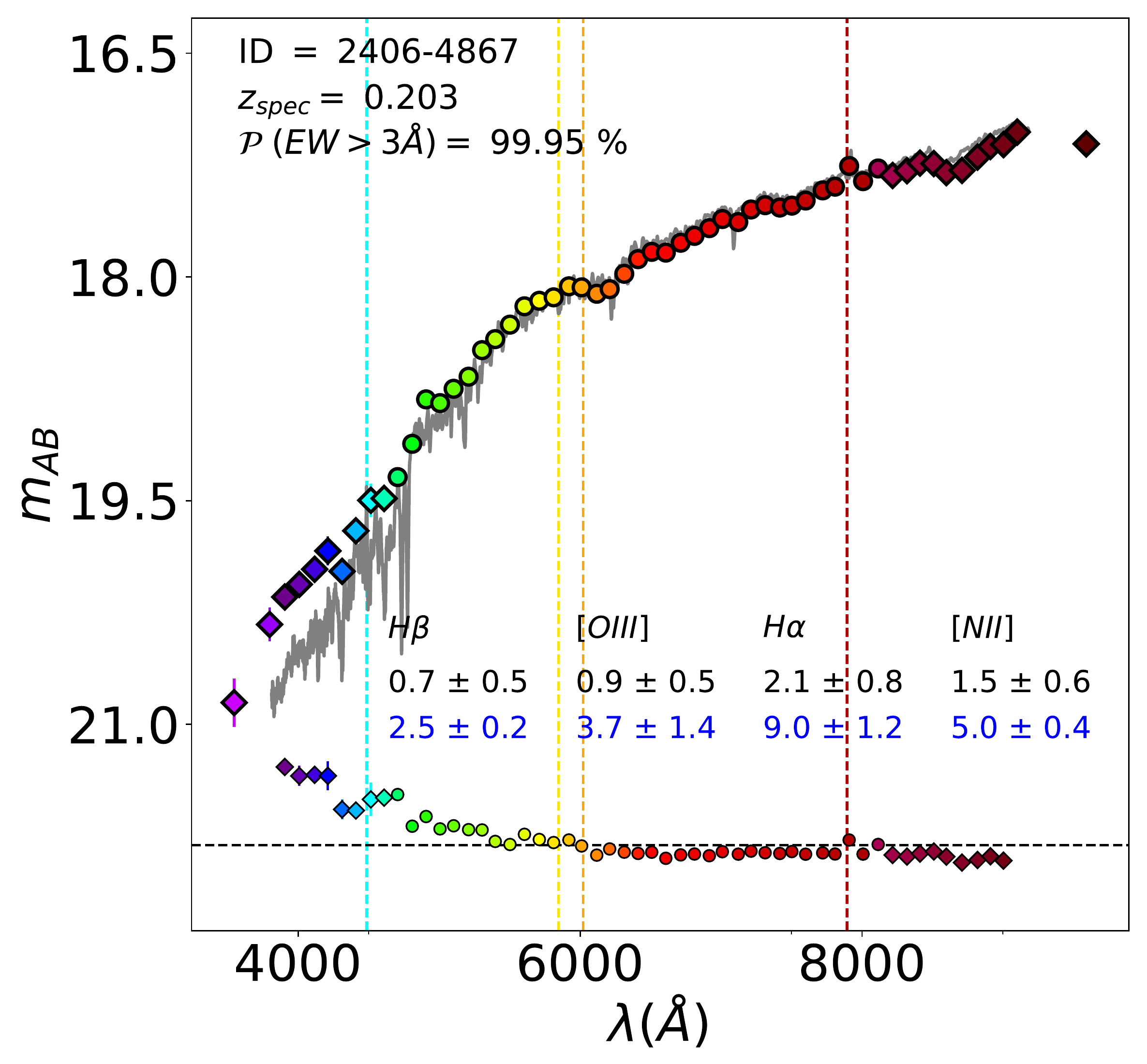}
	\includegraphics[width=6.4cm,height=6cm]{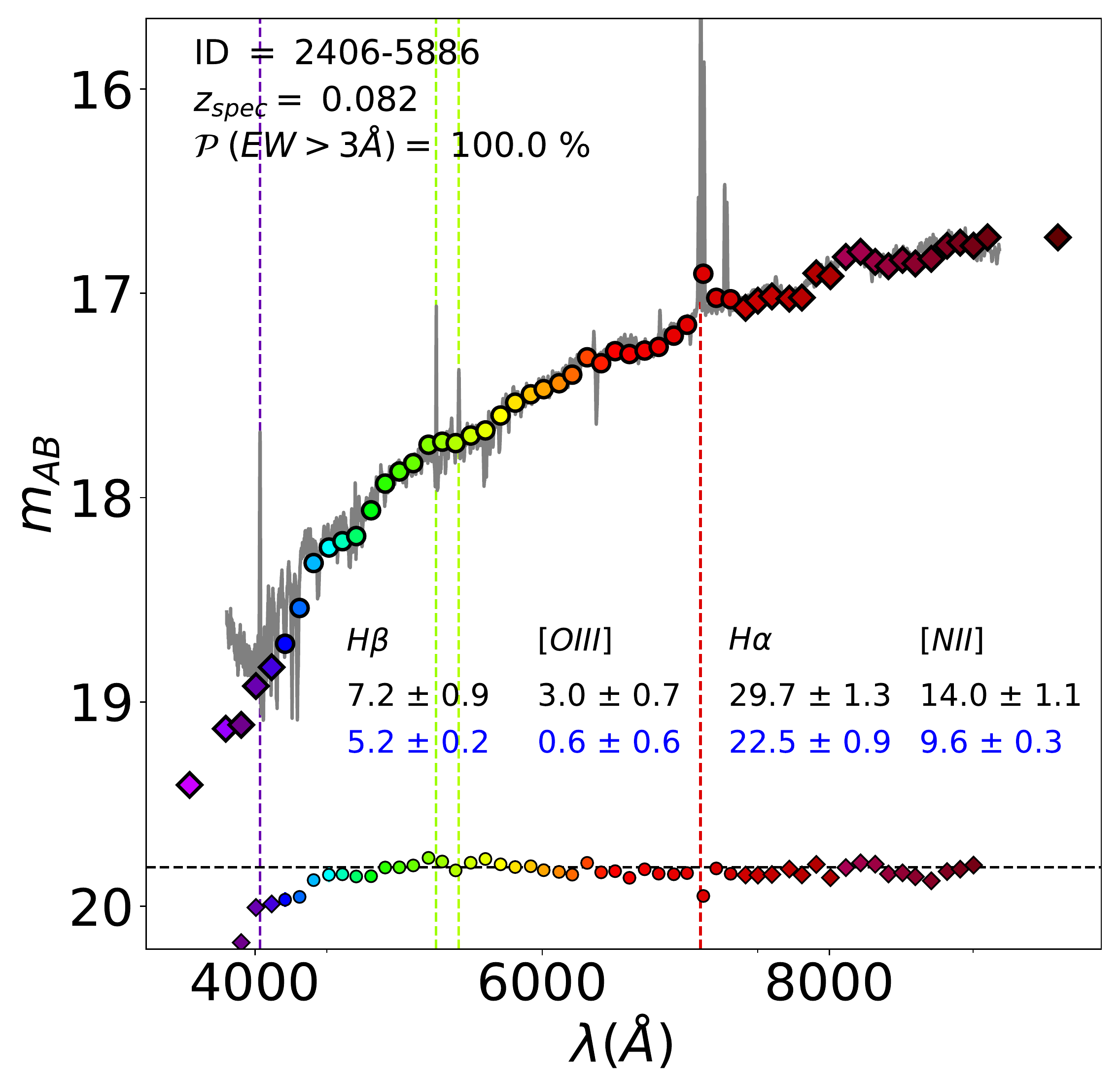}
\caption{\tiny{Examples of J-PAS galaxies in the AEGIS field with SDSS spectrum. The SDSS spectrum is re-scaled to match the rSDSS J-PAS magnitude. Diamonds correspond to the filters not used by the ANN. Blue and black numbers show, respectively, the predictions made by the ANN$_R$ on the EWs and the values measured in the SDSS spectrum. On the top left part of the plot we indicate the J-PAS ID of the object, its redshift and the prediction of the ANN$_{C}$ for $EW_{min} = 3$ \AA}. At the button part we show the difference in magnitude between the synthetic fluxes obtained from SDSS spectra and J-PAS data. Dashed lines mark from left to right the position of \oii, \Hb, \oiii, and \Ha emission lines.}
\label{fig2spec}
\end{figure*}

For this comparison, we select all galaxies observed with SDSS and miniJPAS with redshift below $z \le 0.35$ and minimum average SN of 20 in J-PAS narrow band filters. By a visual inspection we get rid of all QSOs in the sample. We end up with a total of 89 objects. Whenever photometry measurements are lacking or the SN in a particular filter is below $2.5$, we replace it by the best-fit obtained from the stellar population analysis of the galaxy as we discussed in Sec. \ref{miss}. For this comparison we employ \texttt{BaySeAGal} (Amorim in prep), a Bayesian parametric approach which assumes a tau-delayed star formation model for the star formation history. 
\\\\Generally, galaxy properties vary within the galaxy: the distribution of the gas, its temperature and its density, the distribution of interstellar dust or the stellar populations change in function of the position in the galaxy \citep{2015A&A...581A.103G}. Consequently, if the SFR of a galaxy were higher in the outer parts, the galaxy would look younger in the integrated spectrum than in the central part. Similarly, the AGN of a galaxy would not leave the same imprint in the spectrum if the integrated areas covered regions dominated by other ionization mechanisms. Therefore, ideally, one would like to analyse the same region in both surveys, which implies integrating over the same area. However, the aperture corresponding to the $3$ arcsec fiber of SDSS is not sufficiently large to ensure that the Point Spread function (PSF) of J-PAS filter system is not affecting the photometry in the filters where the seeing is worse. For this reason, we make use of the \texttt{MAG\_PSFCOR} photometry which corrects each magnitude individually by considering the light profile of the galaxy and the PSF for each filter \citep{2014MNRAS.441.2891M,2019A&A...622A.178M}. As a consequence, the integrated area varies from galaxy to galaxy, going from 2 to 7 arcsec, and should be taken into account to interpret fairly this comparison. Although the ANN$_R$ only use colors as inputs, we scale the SDSS spectrum to match the rSDSS J-PAS magnitude in each galaxy for a visual inspection. 
\\\\ Figure \ref{figAEGISSDSS} shows the EWs obtained by the ANN$_R$ on J-PAS photometric data (column 1) and on the synthetic J-PAS magnitudes obtained after convolving SDSS spectra with J-PAS filters (column 2) and assuming an average SN of 20. We compare those values with the EWs derived as a result of fitting a Gaussian function to each one of the emission lines in the spectrum (x-axis). We do not include in this comparison the emission lines where EWs are below 1 \AA, which indeed are compatible with zero. The number of galaxies in each row are from top to bottom 57, 37, 64, and 31. We find an excellent agreement when it comes to SDSS synthetic magnitudes, which is in line with the simulations performed with the SDSS dataset. We also find a remarkable correlation in \Ha, \Hb and \nii with J-PAS magnitudes, but we obtain in most of the cases higher values with an increase in the dispersion (see median and MAD in Fig. \ref{figAEGISSDSS}). The agreement is less favourable for \oiii line. Nevertheless, we should bear in mind the limiting number of galaxies used here in order to avoid drawing any conclusion that may not be supported from a statistical point of view. Instead, we consider more appropriated to analyze the origin of these discrepancies by examining visually each object.
\\\\ In Fig. \ref{fig2spec} we show several galaxies analyzed in this comparison. We re-scale the SDSS spectrum to match the rSDSS J-PAS magnitude. We compare the values of the EWs  measured in the SDSS spectrum (black) with the values predicted by the ANN$_R$ (blue) for each one of these galaxies. On the bottom part, we show in each filter the difference between J-PAS data and SDSS synthetic photometry, which certainly can help to shed light on the origin of the discrepancies. 
\\\\ In the first row of Fig. \ref{fig2spec} we display three examples of emission line galaxies where the agreement in most of the EWs is remarkable. Although ANNs are often difficult to interpret, it is evident after a visual inspection that the filters capturing the fluxes of the emission lines are the most relevant in determining the values of the EWs. The excess in the flux of \Ha in galaxy 2243-8838 explains the increase in its EW respect to what it is obtained from a direct measurement in the spectrum or with the synthetic fluxes by means of the ANN$_R$. In the same vein, the drop in the flux observed in the \oiii line in galaxy 2241-12850 clarifies the differences found in the EW. Second order terms include the relation between emission lines (Balmer decrement or recombination lines) and the colors of galaxies. Definitely, the excess in the flux of \Hb in galaxy 2243-9127 does not only increase the value of such line but also contributes to enlarge the EW of \Ha. 
\\\\In the second row of Fig. \ref{fig2spec} we show Early-type galaxies (ETGs) where the differences between J-PAS data and SDSS synthetic fluxes are negligible. The ANN$_C$ estimates very low probability for these galaxies to have any emission line with a EW greater than 3 \AA, which is in agreement with the measurements performed in SDSS spectra. As we discussed in Sect. \ref{subsec:EWdependence} the ANN$_R$ tends to overestimate the EWs in the regime of low emission and consequently a zero level bias appears in these galaxies. Nonetheless, for many of these lines the values are compatible with the uncertainty and never overcome the 3 \AA $ $ limit. 
\\\\Finally, in the third row of Fig. \ref{fig2spec} we focus our attention on galaxies where the fluxes seen by J-PAS and SDSS present evident differences in the blue part of the spectrum. The integrated areas in J-PAS are probably capturing regions with more populations of young stars in 2243-9209 and 2406-4867 galaxies. Such population rises the number of ionising photons and it is responsible of the increase in the EWs of emission lines that we observe. The opposite effect occurs in galaxy 2406-5886, the galaxy looks redder with J-PAS data and the flux in \Ha is less intense. Therefore, the predictions of the ANN$_R$ in the EWs are below the values measured in the SDSS spectrum. 
\\\\ To sum up, despite of the fact that this comparison suffer from several difficulties and it would need many more galaxies to be statistically robust, results are coherent with the simulations presented in Sect. \ref{sec:Simulations} and lay the foundations to better understand and interpret the whole sample of galaxies observed in the AEGIS field that we will analyze in a future work.

 \section{Summary and conclusions}\label{sec:conclusion}
We have developed a new method based on ANNs to measure and detect emission lines in J-PAS up to $z = 0.35$. We can classify galaxies according to the EWs of the emission lines even with high uncertainty in the redshift. This will allow us to better study the density function of emitting-line galaxies in J-PAS. \\\\ With the synthetic photometry of CALIFA, MaNGA or SDSS spectra, we have trained an ANN$_R$ to estimate the EWs of \Ha, \Hb, \nii and \oiii lines. We present two training samples to undertake this task. 
\\\\Firstly, we trained the ANN$_R$ with only synthetic J-spectra from MaNGA and CALIFA surveys and we used SDSS to evaluate the performance of the model. The lack of enough number of AGN-like synthetic J-spectra produces a saturation of  \nii/\Ha and \oiii/\Hb ratios at high values, what compromises the ability of the model to deal with galaxies where the main ionization mechanism is not dominated by star formation processes. Nevertheless, we are able to constrain those ratios within $0.101$ and $0.091$ dex. Furthermore, we are able to reach $0.091$ and $0.087$ dex respectively if one considers only star-forming galaxies. This is a significant improvement in the precision previously. While a method based on the photometry contrast need for an EW of 10 \AA $ $ a SN in the photometry of at least 15.5, the ANN can measure the same EW in \Ha, \Hb, \nii and \oiii lines with a SN in the photometry of  5, 1.5, 3.5 and, 10 respectively.
\\\\Secondly, we trained the ANN$_R$ with SDSS galaxies and we revealed the importance of testing the model with data coming from different surveys. Otherwise, the performance of the model can be overestimated. While the SDSS training set scores very high with SDSS testing sample, the performance worsens when we compare it with MaNGA or CALIFA test sample. 
\\\\Finally, we estimate the EWs of a set of galaxies observed both in SDSS and miniJPAS. We compare the performance of ANN$_{R}$ in the synthetic SDSS fluxes with the performance in the fluxes measured by J-PAS. Despite the difficulty of comparing data from different surveys in equal terms, we reach an overall agreement. We argue that the origin of the discrepancies might be attributed to differences between the integration areas in miniJPAS and SDSS and/or photometry artefacts that appear as a result of the PSF. Many more data would be needed to be conclusive.
\\\\ In this work our model is limited to redshift below $z = 0.35$ in order to ensure \Ha line is measurable with the J-PAS filter system. However, J-PAS will be able to detect galaxies up to $z \sim 1$. Other emission lines such as the $ [OII]{\lambda  \lambda}$ 3726,3729 doublet are visible in the optical range up to redshift $z < 1.6$ and have been used as tracer of star formation in many works \citep{2004AJ....127.2002K,2012MNRAS.420.1926S}. An ultimate version of the model should take into account those facts and build a more sophisticated and complete training sample to be able to overcome the limitations and inaccuracies mentioned so as to fully exploit the potentiality of J-PAS. 
Our main conclusions are summarized below:
\begin{itemize}
  \item The ANN$_C$ can classify galaxies according to the EWs of the emission lines beyond the contrast that one can directly measure with sufficient significance in J-PAS ($\sim 16$ \AA) and also in the case of high uncertainty in the redshift.
  \item The ANN$_R$ trained with the CALMa set can estimate the EWs of \Ha, \Hb, \nii, and \oiii in SDSS galaxies with a relative standard deviation of $8.7 \%$, $14.3 \%$, $15.9 \%$, and $16.4 \%$ respectively. \Ha, \Hb, \nii,, and \oiii lines presents a relative bias of $0.17 \%$, $5.4 \%$, $4.8 \%$, and $-6.4 \%$ respectively. For a SN of 3, the minimum EW measurable in \Ha, \Hb, \oiii and \nii lines are 18, 6, 40 and, 13 \AA. respectively
  \item The \nii/\Ha is constrained within $0.093$ dex and a bias of $-0.019$ dex and the \oiii\Hb ratio with no bias and a dispersion of $0.081$ dex in SDSS galaxies. The $ \ion{O}{3}\ion{N}{2}$ is recovered within $0.114$ dex and a bias of $0.038$ dex.
  \item  We found an overall correlation between miniJPAS and SDSS galaxies in the EW of \Ha, \Hb and, \nii lines. The correlation in the EW of \oiii is less strong. More data will be needed to unveil the origin of such discrepancy. Certainly, the problems associated to the integrated areas are playing an important role.

\end{itemize}

\begin{acknowledgements}
G.M.S., R.G.D, R.G.B., E.P., J.R.M, L.~A.~D.~G.~ and J.M.V. acknowledge support from the State Agency for Research of the Spanish MCIU through the  "Center of Excellence Severo Ochoa" award to the Instituto de Astrofísica de Andalucía (SEV-2017-0709) and the projects AYA2016-77846-P and PID2019-109067-GB100. L.~A.~D.~G.~ acknowledges support from the Ministry of Science and Technology of Taiwan (grant MOST 106-2628-M-001-003-MY3) and from the Academia Sinica (grant AS-IA-107-M01). P.O. Baqui and V. Marra acknowledge support from the Coordenação de Aperfeiçoamento de Pessoal de N\'ivel Superior – Brasil (CAPES) – Finance Code 001. R.A.D. acknowledges support from the CNPq through BP grant 308105/2018-4, and FINEP grants REF. 1217/13 - 01.13.0279.00 and REF 0859/10 - 01.10.0663.00 and also FAPERJ PRONEX grant E-26/110.566/2010 for hardware funding support for the JPAS project through the National Observatory of Brazil and Centro Brasileiro de Pesquisas Físicas. LSJ acknowledges support from Brazilian agencies FAPESP (2019/10923-5) and CNPq (304819/201794). AC acknowledges support from PNPD/CAPES. J.M.V. acknowledges financial support from research projects AYA2016-79724-C4-4-P, PID2019-107408GB-C44 from the Spanish Ministerio de Ciencia e Innovación. The authors acknowledge the following people for providing valuable comments and suggestions on the first draft of this paper: Stravos Akras, Joel Bregman, Salvador Duarte Puertas, Jorge Iglesias, Yolanda Jimenez Teja, Jose Miguel Rodriguez Espinosa, David Sobral and, Adi Zitrin. 
\\\\This research made use of {\tt Python} (http://www.python.org), Numpy \citep{van2011numpy}; of Matplotlib \citep{Hunter:2007}, a suite of open-source Python modules that provides a framework for creating scientific plots and, Astropy, the community-developed core {\tt python} package \citep{astropy:2013, astropy:2018}. Funding for SDSS-III has been provided by the Alfred P. Sloan Foundation, the Participating Institutions, the National Science Foundation, and the U.S. Department of Energy Office of Science. The SDSS-III web site is http://www.sdss3.org. This study makes use of the results based on the Calar Alto Legacy Integral Field Area (CALIFA) survey (http://califa.caha.es/).This project made use of the MaNGA-Pipe3D dataproducts. We thank the IA-UNAM MaNGA team for creating this catalogue, and the ConaCyt-180125 project for supporting them. 
\\\\ Funding for the J-PAS Project has been provided by the Governments of España and Aragón though the Fondo de Inversión de Teruel, European FEDER funding and the MINECO and by the Brazilian agencies FINEP, FAPESP, FAPERJ and by the National Observatory of Brazil. Based on observations made with the JST/T250 telescope and PathFinder camera for the miniJPAS project at the Observatorio Astrof\'{\i}sico de Javalambre (OAJ), in Teruel, owned, managed, and operated by the Centro de Estudios de F\'{\i}sica del  Cosmos de Arag\'on (CEFCA). We acknowledge the OAJ Data Processing and Archiving Unit (UPAD) for reducing and calibrating the OAJ data used in this work.Funding for OAJ, UPAD, and CEFCA has been provided by the Governments of Spain and Arag\'on through the Fondo de Inversiones de Teruel; the Arag\'on Government through the Research Groups E96, E103, and E16\_17R; the Spanish Ministry of Science, Innovation and Universities (MCIU/AEI/FEDER, UE) with grant PGC2018-097585-B-C21; the Spanish Ministry of Economy and Competitiveness (MINECO/FEDER, UE) under AYA2015-66211-C2-1-P, AYA2015-66211-C2-2, AYA2012-30789, and ICTS-2009-14; and European FEDER funding (FCDD10-4E-867, FCDD13-4E-2685).
\end{acknowledgements}

\appendix
\section{SDSS training set}\label{appSDSS}
\begin{figure*}
    \includegraphics[width= 0.94\linewidth]{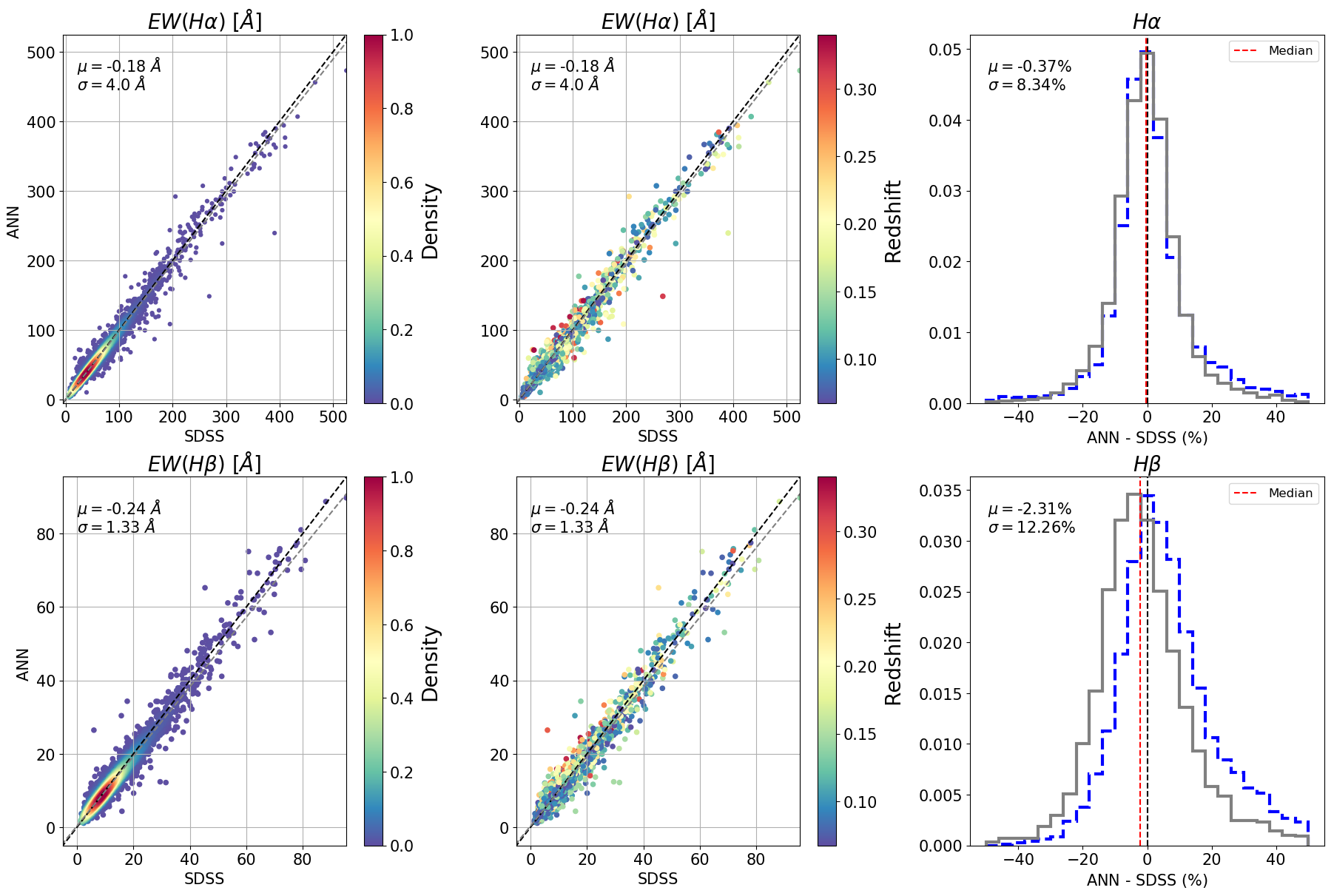}
    \includegraphics[width= 0.94\linewidth]{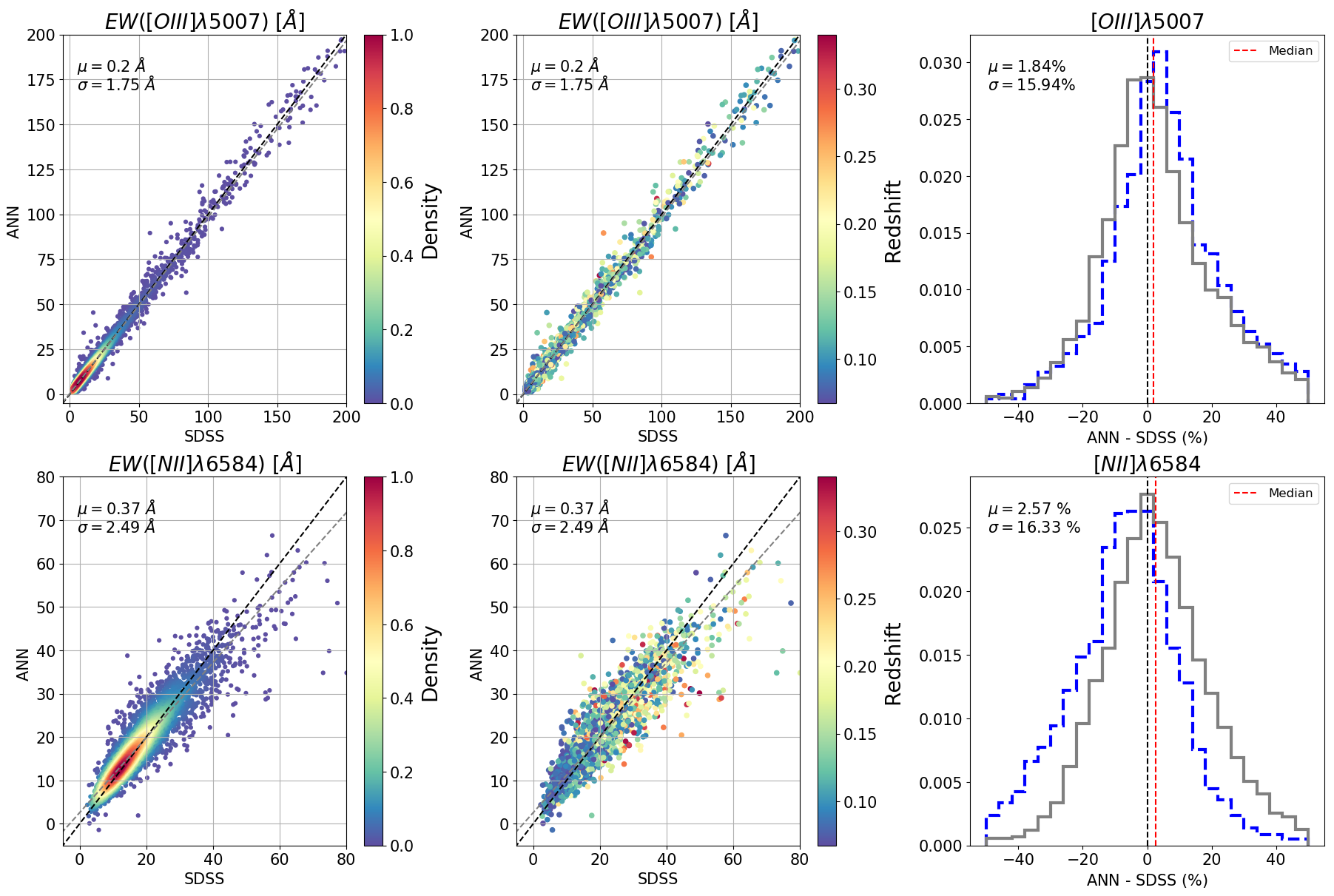}
 \caption{\tiny{EWs of \Ha, \Hb, \nii and \oiii predicted by the ANN$_R$ compared to SDSS testing sample.The ANN$_R$ is trained with the SDSS training set. The color-code represents the probability density function defined by a Gaussian kernel  (right panel) and the redshift of the objects (left panel). The histograms in grey are normalized to one and show the relative difference between both values. The histograms in blue are the ones in Fig. \ref{fig:lines} and are shown for a visual comparison.} Black and blue numbers are the median and the median absolute deviation of the difference. Black and grey dashed lines on the left are lines with slope one and the best linear fit respectively. We perform a sigma clipping fit with $\sigma = 3$ to exclude outliers. Red dashed line represents the median.}
\label{linesSDSS}
\end{figure*}

\begin{figure*}
    \includegraphics[width=\linewidth]{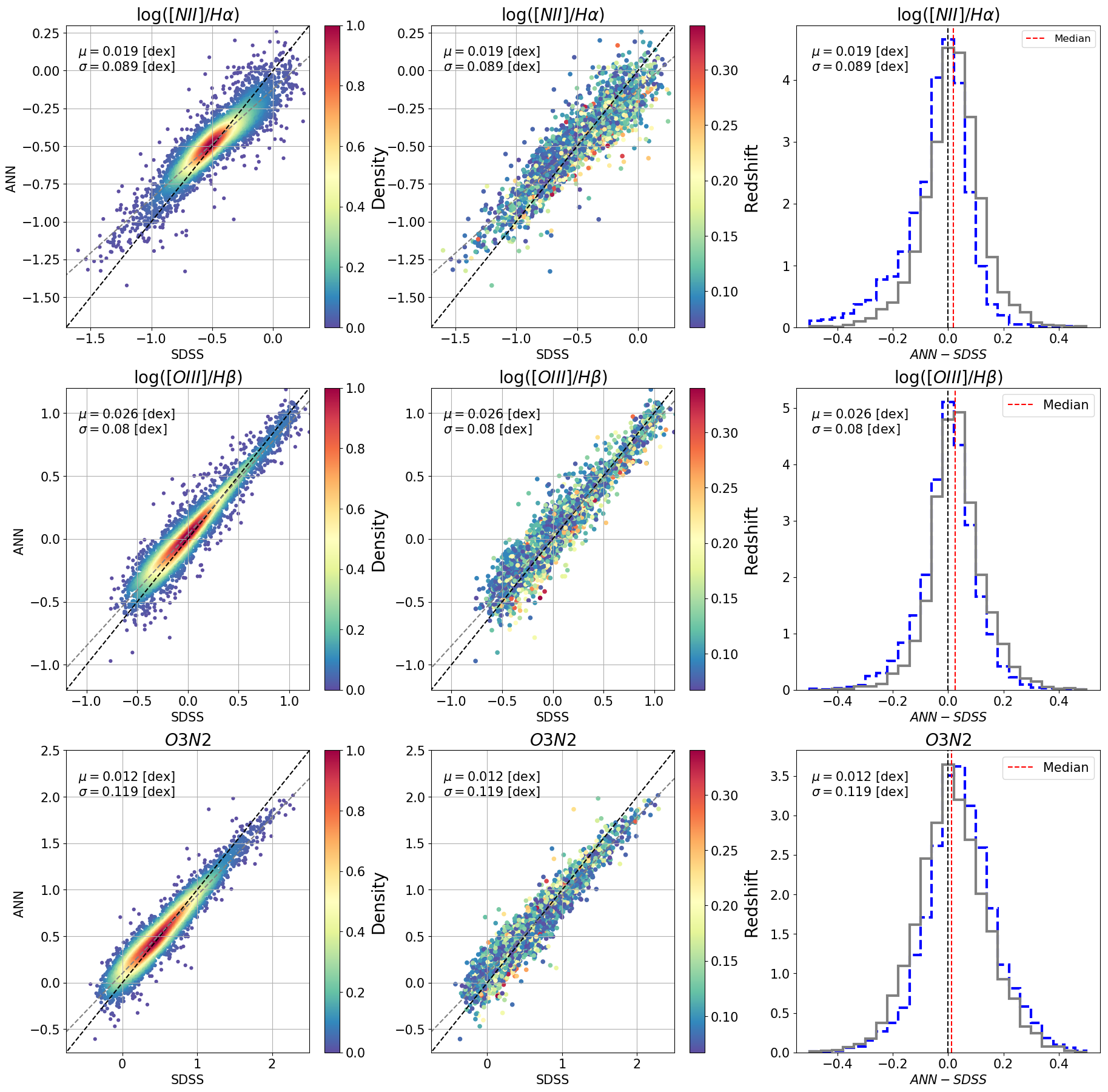}
 \caption{\tiny{Comparison between \nii/\Ha, \oiii/\Hb and $ \ion{O}{3}\ion{N}{2}$ ratios estimated by the ANN$_R$ and SDSS testing sample. The ANN$_R$ is trained with the SDSS training set Same scheme of Fig. \ref{linesSDSS}.} }
\label{ratiosSDSS}
\end{figure*}

\begin{figure*}
    \includegraphics[width=\linewidth]{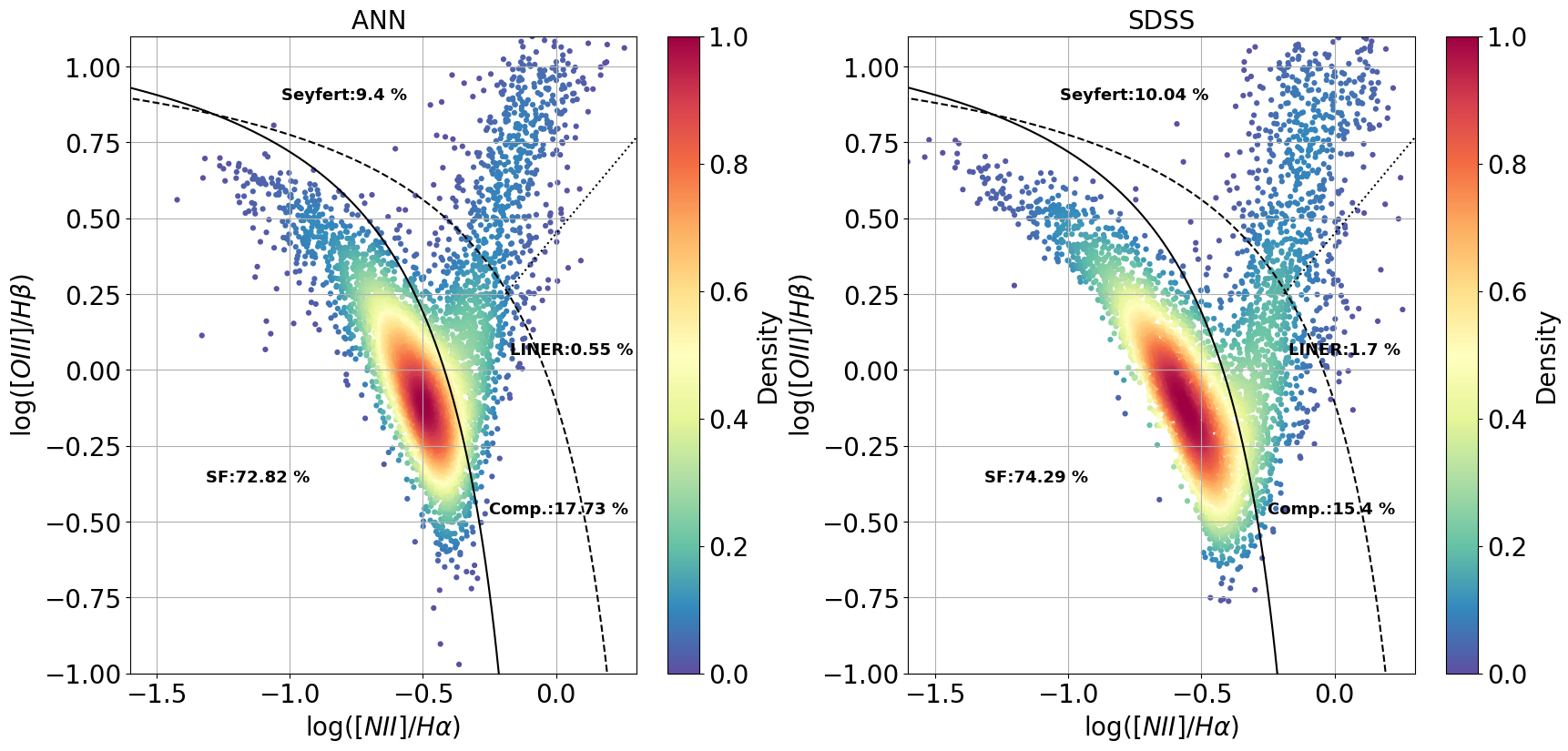}
 \caption{\tiny{ BPT diagram obtained with the ANN$_R$ and SDSS MPA-JHU DR8 catalog where the color-code indicates the density of points. The ANN$_R$ is trained with the SDSS training set. The solid (ka03), dashed (Ke01) and dotted lines (S07) define the regions for the four main ionization mechanism of galaxies. The percentage for each group is shown in black.}}
\label{fig:BPTSDSS}
\end{figure*}

In this section we show how the SDSS training set scores in the SDSS testing sample. This represents the ideal situation where the testing set is included within the parameter space of the training set. In other words, the testing sample is a subset of the training set and consequently the only uncertainties found in the targets variables (EWs) area associated to the capability of the ANN$_R$ algorithm to decode the information provided by the inputs (J-spectrum). Nonetheless, we cannot infer from that the actual potential of  the ANN$_R$ to predict in J-PAS data. As we discussed in the main body of this paper, here lies the reason why the ANN$_R$ must be tested with data with different observational setup and calibrations.
\\\\In Fig. \ref{linesSDSS} we plot the EWs predicted by the ANN$_R$ versus the EWs provided by the SDSS testing sample from the MPA-JHU DR8 catalog. This plot follows the same scheme of Fig. \ref{fig:lines}.  As happened with the CALMa  training set, we constrain better the EW of \Ha followed by \Hb, \oiii and \nii. However, the \nii line is recovered with no bias and it does not saturate at high values. 
\\\\ In Fig. \ref{ratiosSDSS} we show the comparison between the logarithmic ratios of \nii/\Ha, \oiii/\Hb and $ \ion{O}{3}\ion{N}{2}$ in a similar way as we did in Fig. \ref{fig:ratios}. The \nii/\Ha ratio is predicted within $0.089$ dex and a bias of $0.019$ dex and the \oiii/\Hb ratio within $0.08$ dex and a bias of $0.026$ dex. As a result, the $\ion{O}{3}\ion{N}{2}$ is recovered within $0.119$ dex and a bias of $0.012$. 
\\\\Finally, we show in Fig. \ref{fig:BPTSDSS} a comparison of the BPT diagram recovered by the ANN$_R$ (left plot) and the one obtained from the SDSS testing sample (right plot) following once again the same scheme of Fig. \ref{fig:BPT}. The similarity between those diagrams is remarkable. We are not only able to recover properly the SF-wing but also the AGN branch, obtaining similar percentages of galaxies in all the regions. 

\bibliographystyle{aa}
\bibliography{aa}
\end{document}